\documentclass[preprint,prd,aps,showpacs,showkeys,nofootinbib]{revtex4}
\usepackage{graphicx}
\begin{document}


\title{Higgs boson decays with lepton flavor violation in the $\emph{B-L}$ symmetric SSM}

\author{Ze-Ning Zhang$^{1}$\footnote{zn\_zhang\_zn@163.com},
Hai-Bin Zhang$^{1,2}$\footnote{Corresponding author.\\hbzhang@hbu.edu.cn},
Jin-Lei Yang$^{1,3}$,
Shu-Min Zhao$^{1,2}$,
and Tai-Fu Feng$^{1,2,4}$\footnote{fengtf@hbu.edu.cn}}

\affiliation{$^1$Department of Physics, Hebei University, Baoding, 071002, China\\
$^2$Key Laboratory of High-precision Computation and Application of Quantum Field Theory of Hebei Province, Baoding, 071002, China\\
$^3$Institute of theoretical Physics, Chinese Academy of Sciences, Beijing, 100190, China\\
$^4$College of Physics, Chongqing University, Chongqing, 400044, China}

\begin{abstract}
Recently, the ATLAS and CMS Collaborations reported the latest experimental upper limits on the branching ratios of the lepton flavor violating (LFV) 125 GeV Higgs boson decays, $h\rightarrow e\mu$, $h\rightarrow e\tau$, and $h\rightarrow \mu\tau$. In this paper, we mainly investigate the LFV Higgs boson decays $h\rightarrow e\mu$, $h\rightarrow e\tau$, and $h\rightarrow \mu\tau$ in the minimal supersymmetric extension of the Standard Model with local $\emph{B-L}$ gauge symmetry. At the same time, the corresponding constraints from the LFV rare decays $\mu\rightarrow e\gamma$, $\tau\rightarrow e\gamma$, $\tau\rightarrow \mu\gamma$, and muon $(g-2)$ are considered to analyze the numerical results.

\end{abstract}

\keywords{Supersymmetry, Higgs Boson Decay, Lepton Flavor Violation}
\pacs{12.60.Jv, 14.80.Da, 11.30.Fs}

\maketitle

\section{Introduction\label{sec1}}
Recently, a search for lepton flavor violating (LFV) decays of the 125 GeV Higgs boson in the $e\mu$, $e\tau$, and $\mu\tau$ decay modes is presented. According to the latest experimental data provided by ATLAS and CMS, the observed upper limits on the LFV branching ratio of $h\rightarrow e\mu$, $h\rightarrow e\tau$, and $h\rightarrow \mu\tau$ at 95\% confidence level (C.L.) are~\cite{TAC2020,TAC2020-1,TCC2018}
\begin{eqnarray}
&&{\rm{Br}}(h\rightarrow e\mu)<6.2\times 10^{-5},\label{a1}\\
&&{\rm{Br}}(h\rightarrow e\tau)<4.7\times 10^{-3},\label{a2}\\
&&{\rm{Br}}(h\rightarrow \mu \tau)<2.5\times 10^{-3}.\label{a3}
\end{eqnarray}
It is interpreted as a signal, $e\mu$ means the final state consisting of $\bar{e} \mu$ and $e\bar{\mu}$, $e\tau$ means the final state consisting of $\bar{e} \tau$ and $e\bar{\tau}$, and $\mu \tau$ means the final state consisting of $\bar{\mu} \tau$ and $\mu\bar{\tau}$. The LFV Higgs boson decays are forbidden in the Standard Model (SM)~\cite{Harnik}. But they can easily occur in new physics (NP) models beyond the SM, for instance, the supersymmetric models~\cite{LFVHD-Br1,LFVHD2,LFVHD2-0,LFVHD2-1,LFVHD2-2,LFVHD-Br2,LFVHD2-3,LFVHD2-4,LFVHD3,HTSYF, ref-zhang-LFV, ref-zhang,ref-zhang1,50}, the composite Higgs model~\cite{LFVHD4,LFVHD5}, the two Higgs doublet Model~\cite{add19,add20,add21,add9,add1,add2}, and others~\cite{add10,add12,LFVHD46,add15,LFVHD47,LFVHD60,LFVHD49,LFVHD6,LFVHD10,LFVHD7,LFVHD8,LFVHD9,
LFVHD11,LFVHD12,LFVHD50,LFVHD51,LFVHD62,LFVHD13,LFVHD14,LFVHD15,add24,add28,LFVHD15-1,LFVHD48}. Due to the running of the Large Hadron Collider~(LHC), LFV Higgs decays have recently been discussed within various theoretical frameworks~\cite{LFVHD55,LFVHD52,LFVHD20,LFVHD21,LFVHD22,LFVHD23,LFVHD24,LFVHD25,LFVHD26,LFVHD27,LFVHD53,LFVHD57,LFVHD28,
LFVHD29,LFVHD30,LFVHD31,LFVHD32,LFVHD33,LFVHD34,LFVHD36,LFVHD42-1,LFVHD56,LFVHD63,LFVHD64,LFVHD65,LFVHD66,add29,LFVHD35,
LFVHD37,LFVHD38,LFVHD39,LFVHD40,LFVHD41,LFVHD42,LFVHD42-2,LFVHD43,LFVHD44,LFVHD59,LFVHD67,LFVHD68,LFVHD69,LFVHD70,LFVHD71,LFVHD71-1,
LFVHD72,LFVHD73,LFVHD74,LFVHD75,LFVHD76,LFVHD77,LFVHD78,LFVHD79,LFVHD79-1,LFVHD80,LFVHD81,LFVHD81-1,LFVHD82,LFVHD83,LFVHD84,LFVHD85,LFVHD86,
LFVHD88,LFVHD89,LFVHD90,LFVHD91,add74,add76,add80,add81,add82,add83,add84,add85,add86,add87,add88,add90,add91,
Nguyen:2020ehj,Hou:2020tgl,Hong:2020qxc,Hung:2021}. Lepton flavor violation is to provide a window into the search for NP. In this work, we will investigate the LFV Higgs boson decays $h\rightarrow e\mu$, $h\rightarrow e\tau$, and $h\rightarrow \mu\tau$ in the minimal supersymmetric extension of the Standard Model with local $\emph{B-L}$  gauge symmetry (B-LSSM)~\cite{5,6,46,47,48,49,B-L1,B-L2}.

The gauge symmetry group of the B-LSSM extends that of the minimum supersymmetric Standard Model (MSSM)~\cite{MSSM,MSSM1,MSSM2,MSSM3,MSSM4} to $SU(3)_C\otimes SU(2)_L\otimes U(1)_Y\otimes U(1)_{B-L}$, where $B$ stands for the baryon number and $L$ for the lepton number. The conservation of $B$ and $L$ is normally connected to R-parity. R-parity can be written as $R=(-1)^{3(B-L)+2S}$, where $B$, $L$, and $S$ are baryon number, lepton number, and spin, respectively. The invariance under $U(1)_{B-L}$ gauge group imposes the R-parity conservation which is assumed in the MSSM to avoid proton decay. And R-parity conservation can be maintained if $U(1)_{B-L}$ symmetry is broken spontaneously while ensuring that the scalars breaking the symmetry have even charges under the B-LSSM~\cite{C.S.A}. Furthermore, the B-LSSM can provide much more candidates for the dark matter comparing that with the MSSM, for example, new neutralinos corresponding to the gauginos of $U(1)_{B-L}$, additional Higgs singlets and sneutrinos~\cite{16,1616,DelleRose:2017ukx,DelleRose:2017uas}. In the B-LSSM, magnetic and electric dipole moments of leptons and quarks have been analyzed~\cite{50,MDM-1,MDM-2,MDM-3}.

In our previous work, we studied the lepton flavor violating decays $l_j^-\rightarrow l_i^-\gamma$ and $l_j^-\rightarrow l_i^-l_i^-l_i^+$ in the B-LSSM~\cite{50}.  The numerical results show that the present experimental limits for the branching ratio of $l_j^-\rightarrow l_i^-\gamma$ constrain the parameter  space of the  B-LSSM  most strictly. In this work, considering the constraint of the present experimental limits on the branching ratio of $l_j^-\rightarrow l_i^-\gamma$, we give the influence of slepton flavor mixing parameters for the branching ratio of $h\rightarrow l_il_j$ and $l_j^-\rightarrow l_i^-\gamma$ in the B-LSSM. The upper limits on the LFV branching ratio of $\mu\rightarrow e\gamma$, $\tau\rightarrow e\gamma$, and $\tau\rightarrow \mu\gamma$ at 90\% C.L. are now given by~\cite{AMB44,BATBC45,KNHOYU46}
\begin{eqnarray}
&&{\rm{Br}}(\mu\rightarrow e\gamma)<4.2\times10^{-13},\label{a4}\\
&&{\rm{Br}}(\tau\rightarrow e\gamma)<3.3\times10^{-8},\label{a5}\\
&&{\rm{Br}}(\tau\rightarrow \mu\gamma)<4.4\times10^{-8}.\label{a6}
\end{eqnarray}

The paper is organized as follows. In Sec. II, we mainly introduce the B-LSSM including its superpotential and the general soft breaking terms. In Sec. III, we give an analytic expression for the branching ratio of the 125 GeV Higgs boson decays with LFV in the B-LSSM. In Sec. IV, we give the numerical analysis, and the summary is given in Sec. V. Finally, some tedious formulas are collected in the Appendices.

\section{The B-LSSM\label{sec2}}
Extended the superfields of the MSSM, the B-LSSM~\cite{46,47,48,49,B-L1,B-L2} adds two singlet Higgs fields $\hat{\eta}_{1}$ and $\hat{\eta}_{2}$ and three generations of right-handed neutrinos  $\hat \nu^c_i$. The quantum numbers of the  gauge symmetry group for the chiral superfields in the B-LSSM can be seen in Table~\ref{tab1}. Then, the superpotential in the model can be given by
\begin{eqnarray}
&&W=Y_{u,ij}\hat{Q_i}\hat{H_2}\hat{U_j^c}+\mu \hat{H_1} \hat{H_2}-Y_{d,ij} \hat{Q_i} \hat{H_1} \hat{D_j^c}
-Y_{e,ij} \hat{L_i} \hat{H_1} \hat{E_j^c}\nonumber\\
&&\;\;\;\;\;\;\;\;\;+Y_{\nu, ij}\hat{L_i}\hat{H_2}\hat{\nu}^c_j-\mu' \hat{\eta}_1 \hat{\eta}_2
+Y_{x, ij} \hat{\nu}_i^c \hat{\eta}_1 \hat{\nu}_j^c,
\end{eqnarray}
where $\hat H_1^T = \Big( {\hat H_1^1 ,\hat H_1^2} \Big)$, $\hat H_2^T = \Big( {\hat H_2^1 ,\hat H_2^2} \Big)$, $\hat Q_i^T = \Big( {\hat u_i ,\hat d_i} \Big)$, and $\hat L_i^T = \Big( {\hat \nu_i , \hat e_i} \Big)$ are SU(2) doublet superfields. $\hat U_i^c $, $ \hat D_i^c$, and $\hat E_i^c $ represent up-type quark, down-type quark, and charged lepton singlet superfields, respectively. The dimensionless Yukawa coupling parameter $Y$ is a 3$\times$3 matrix. $i,j=1,2,3$ are the generation indices.  The summation convention is implied on repeated indices.

\begin{table*}
\begin{tabular*}{\textwidth}{@{\extracolsep{\fill}}lllll@{}}
\hline
Superfield & Spin0 & Spin{1/2} & Generations & $SU(3)_C\otimes SU(2)_L\otimes U(1)_Y\otimes U(1)_{B-L}$\\
\hline
$\hat{Q}_i$ & $\widetilde{Q}_i$ & $Q_i$ & 3 & (3, 2, 1/6, 1/6)\\
$\hat{D}^c_i$ & $\widetilde D^c_i$ & $D^c_i$ & 3 & ($\bar{3}$, 1, 1/3, {$\rm{-1/6}$})\\
$\hat{U}^c_i$ & $\widetilde U^c_i$ & $U^c_i$ & 3 & ($\bar{3}$, 1, {$\rm{-2/3}$}, {$\rm{-1/6}$})\\
$\hat{L}_i$ & $\widetilde L_i$ & $L_i$ & 3 & (1, 2, {$\rm{-1/2}$}, {$\rm{-1/2}$})\\
$\hat{E}^c_i$ & $\widetilde E^c_i$ & $E^c_i$ & 3 & (1, 1, 1, {1/2})\\
$\hat{\nu}^c_i$ & $\widetilde{\nu}^c_i$ & $\nu^c_i$ & 3 & (1, 1, 0, {1/2})\\
$\hat{H}_1$ & $H_1$ &$\widetilde H_1$ &1 & (1, 2, {$\rm{-1/2}$}, 0)\\
$\hat{H}_2$ & $H_2$ &$\widetilde H_2$ &1 & (1, 2, 1/2, 0)\\
$\hat{\eta}_1$ & $\eta_1$ &$\widetilde{\eta}_1$ &1 & (1, 1, 0, $\rm{-1}$)\\
$\hat{\eta}_2$ & $\eta_2$ &$\widetilde{\eta}_2$ &1 & (1, 1, 0, 1)\\
\hline
\end{tabular*}
\caption{Chiral superfields and their quantum numbers.}
\label{tab1}
\end{table*}

Correspondingly, the soft breaking terms of the B-LSSM are generally given as
\begin{eqnarray}
&&\mathcal{L}_{soft}=-
m_{\tilde{q},ij}^2\tilde{Q}_i^*\tilde{Q}_j-m_{\tilde{u},ij}^2(\tilde{u}_i^c)^*\tilde{u}_j^c
-m_{\tilde{d},ij}^2(\tilde{d}_i^c)^*\tilde{d}_j^c-m_{\tilde{L},ij}^2\tilde{L}_i^*\tilde{L}_j
-m_{\tilde{e},ij}^2(\tilde{e}_i^c)^*\tilde{e}_j^c\nonumber\\
&&\hspace{1.4cm}
-m_{\tilde{\nu},ij}^2(\tilde{\nu}_i^c)^* \tilde{\nu}_j^c-m_{\tilde{\eta}_1}^2 |\tilde{\eta}_1|^2-m_{\tilde{\eta}_2}^2 |\tilde{\eta}_2|^2
-m_{H_1}^2|H_1|^2-m_{H_2}^2|H_2|^2\nonumber\\
&&\hspace{1.4cm}
+\Big[-B_\mu H_1H_2 -B_{\mu^{'}}\tilde{\eta}_1 \tilde{\eta}_2 +T_{u,ij}\tilde{Q}_i\tilde{u}_j^cH_2+T_{d,ij}\tilde{Q}_i\tilde{d}_j^cH_1+T_{e,ij}\tilde{L}_i\tilde{e}_j^cH_1\nonumber\\
&&\hspace{1.4cm}
+T_{\nu}^{ij} H_2 \tilde{\nu}_i^c \tilde{L}_j+T_x^{ij} \tilde{\eta}_1 \tilde{\nu}_i^c \tilde{\nu}_j^c-\frac{1}{2}(M_1\tilde{\lambda}_{B} \tilde{\lambda}_{B}+M_2\tilde{\lambda}_{W} \tilde{\lambda}_{W}+M_3\tilde{\lambda}_{g} \tilde{\lambda}_{g}\nonumber\\
&&\hspace{1.4cm}
+2M_{BB^{'}}\tilde{\lambda}_{B^{'}}\tilde{\lambda}_{B}+M_{B^{'}}\tilde{\lambda}_{B^{'}} \tilde{\lambda}_{B^{'}})+h.c.\Big].
\end{eqnarray}
Here the soft breaking terms mainly contain the mass square terms of squarks, sleptons, sneutrinos, and Higgs bosons, the trilinear scalar coupling terms, and the Majorana mass terms. $\lambda_B$ and $\lambda_B'$ denote the gauginos of $U(1)_Y$ and $U(1)_{B-L}$, respectively. The $SU(2)_L\otimes U(1)_Y\otimes U(1)_{B-L}$ gauge groups break to  $SU(3)_c\otimes U(1)_{em}$ as the Higgs fields receive vacuum expectation values (VEVs),
\begin{eqnarray}
&&H_1^1=\frac{1}{\sqrt2}(v_1+{\rm Re}H_1^1+i{\rm Im}H_1^1),
\qquad\; H_2^2=\frac{1}{\sqrt2}(v_2+{\rm Re}H_2^2+i{\rm Im}H_2^2),\nonumber\\
&&\tilde{\eta}_1=\frac{1}{\sqrt2}(u_1+{\rm Re}\tilde{\eta}_1+i{\rm Im}\tilde{\eta}_1),
\qquad\;\quad\;\tilde{\eta}_2=\frac{1}{\sqrt2}(u_2+{\rm Re}\tilde{\eta}_2+i{\rm Im}\tilde{\eta}_2)\;.
\end{eqnarray}
Here, we define $u^2=u_1^2+u_2^2,\; v^2=v_1^2+v_2^2$, and $\tan\beta^{'}=\frac{u_2}{u_1}$ in analogy to the ratio of the MSSM VEVs ($\tan\beta=\frac{v_2}{v_1}$).

The presence of two Abelian gauge groups gives rise to a new effect absent in any model with only one Abelian gauge group: the gauge kinetic mixing. This mixing couples the $\emph{B-L}$ sector to the MSSM sector, and even if it is set to zero at the grand unification theory scale, it can be induced via renormalization group equations~\cite{RGE1,RGE2,RGE3,RGE4,RGE5,RGE6,RGE7}. In practice, it turns out that it is easier to work with noncanonical covariant derivatives instead of off-diagonal field-strength tensors. However, both approaches are equivalent~\cite{R.F}. Hence, in the following, we consider covariant derivatives of the form
\begin{eqnarray}
D_{\mu}=\partial_\mu-iK^TGA,
\end{eqnarray}
where $ K^T = \Big( {Y ,B-L} \Big)$, $ A^T =\Big({A_\mu^{'Y},A_\mu^{'BL}} \Big)$, $K$ is a vector that contains $Y$ and $\emph{B-L}$ corresponding to hypercharge and $\emph{B-L}$ charge, and $A_\mu^{'Y}$ and $A_\mu^{'BL}$ are the gauge fields. $G$ is the gauge coupling matrix as follows:
\begin{eqnarray}
G=\left(\begin{array}{cc}g_{_Y},&g_{_{YB}}^{'}\\g_{_{BY}}^{'},&g_{_{B-L}}\end{array}\right).
\end{eqnarray}
As long as the two Abelian gauge groups are unbroken, we have the freedom to perform a change of basis by suitable rotation, and $R$ is the proper way to do it.
\begin{eqnarray}
&&\left(\begin{array}{cc}g_{_Y},&g_{_{YB}}^{'}\\g_{_{BY}}^{'},&g_{_{B-L}}\end{array}\right)
R^T=\left(\begin{array}{cc}g_{_1},&g_{_{YB}}\\0,&g_{_{B}}\end{array}\right)\;.
\end{eqnarray}
Here $g_1$ corresponds to the measured hypercharge coupling, which is modified in the B-LSSM and given together with $g_B$ and $g_{YB}$ in Ref.\cite{BLSSM1}. Next, we can redefine the $U(1)$ gauge fields.
\begin{eqnarray}
&&R\left(\begin{array}{c}A_{_\mu}^{\prime Y} \\ A_{_\mu}^{\prime BL}\end{array}\right)
=\left(\begin{array}{c}A_{_\mu}^{Y} \\ A_{_\mu}^{BL}\end{array}\right)\;.
\end{eqnarray}

Since we will be dealing with the decay of Higgs bosons later, we want to give the mass square matrix of scalar Higgs bosons by gauge kinetic mixing as follows:
\begin{eqnarray}
&&M_h^2=u^2\times\nonumber\\
&&\left(\begin{array}{*{20}{c}}
{\frac{1}{4}\frac{g^2 x^2}{1+\tan\beta^2}+n^2\tan\beta}&{-\frac{1}{4}g^2\frac{x^2\tan\beta}{1+\tan^2\beta}}-n^2&
{\frac{1}{2}g_{_B}g_{_{YB}}\frac{x}{T}}&
{-\frac{1}{2}g_{_B}g_{_{YB}}\frac{x\tan\beta'}{T}}\\ [6pt]
{-\frac{1}{4}g^2\frac{ x^2\tan\beta}{1+\tan^2\beta}}-n^2&{\frac{1}{4}\frac{g^2\tan^2\beta x^2}{1+\tan\beta^2}+\frac{n^2}{\tan\beta}}&
{\frac{1}{2}g_{_B}g_{_{YB}}\frac{x\tan\beta}{T}}&{\frac{1}{2}g_{_B}g_{_{YB}}\frac{x\tan\beta\tan\beta'}{T}}\\ [6pt]
{\frac{1}{2}g_{_B}g_{_{YB}}\frac{x}{T}}&{\frac{1}{2}g_{_B}g_{_{YB}}\frac{x\tan\beta}{T}}&{\frac{g_{_B}^2}{1+\tan^2\beta'}+\tan\beta'N^2}&
{-g_{_B}^2\frac{\tan\beta'}{1+\tan^2\beta'}-N^2}\\ [6pt]
{-\frac{1}{2}g_{_B}g_{_{YB}}\frac{x\tan\beta'}{T}}&{\frac{1}{2}g_{_B}g_{_{YB}}\frac{x\tan\beta\tan\beta'}{T}}&
{-g_{_B}^2\frac{\tan\beta'}{1+\tan^2\beta^{'}}-N^2}&{g_{_B}^2\frac{\tan^2\beta'}{1+\tan^2\beta'}+\frac{N^2}{\tan\beta'}}
\end{array}\right),
\end{eqnarray}
where $g^2=g_{_1}^2+g_{_2}^2+g_{_{YB}}^2$, $T=\sqrt{1+\tan^2\beta}\sqrt{1+\tan^2\beta'}$,
$n^2=\frac{{\rm Re}B\mu}{u^2}$ and $N^2=\frac{{\rm Re}B\mu^{'}}{u^2}$, respectively. The mass matrices  in the B-LSSM can be obtained with the help of SARAH~\cite{164,165,166,167,168}.

The mass of the SM-like Higgs boson in the B-LSSM can be written as
\begin{eqnarray}
&&m_h=\sqrt{(m_{h_1}^0)^2+\Delta m_h^2},
\label{15}
\end{eqnarray}
where $m_{h_1}^0$ is the lightest tree-level Higgs boson mass. $\Delta m_h^2$ is the radiative correction for which an approximate expression including two-loop leading-log radiative corrections can be given as ~\cite{HiggsC1,HiggsC2,HiggsC3}
\begin{eqnarray}
&&\Delta m_h^2=\frac{3m_t^4}{4\pi^2 v^2}\Big[\Big(\tilde{t}+\frac{1}{2}\tilde{X}_t\Big)+\frac{1}{16\pi^2}\Big(\frac{3m_t^2}{2v^2}-32\pi\alpha_3\Big)\Big(\tilde{t}^2
+\tilde{X}_t \tilde{t}\Big)\Big],\nonumber\\
&&\tilde{t}=\log\frac{M_S^2}{m_t^2},\qquad\;\tilde{X}_t=\frac{2\tilde{A}_t^2}{M_S^2}\Big(1-\frac{\tilde{A}_t^2}{12M_S^2}\Big),
\label{eq16}
\end{eqnarray}
where $\alpha_3$ is the strong coupling constant, $M_S=\sqrt{m_{\tilde t_1}m_{\tilde t_2}}$ with $m_{\tilde t_{1,2}}$ denoting the stop masses, $\tilde{A}_t=A_t-\mu \cot\beta$ with $A_t=T_{u,33}$ being the trilinear Higgs stop coupling and $\mu$ denoting the Higgsino mass parameter.

The Higgs boson's existence was confirmed by the ATLAS and CMS Collaborations~\cite{ATLAS,CMS} at CERN in 2012 based on colliding experiments at the LHC. The measured mass of the SM-like Higgs boson now is~\cite{PDGPA}
\begin{eqnarray}
m_h=125.10\pm 0.14\: {\rm{GeV}}.
\label{17}
\end{eqnarray}
In the subsequent numerical analysis, we use Eqs.~(\ref{15}) and (\ref{eq16}) to calculate the mass of the SM-like Higgs boson and select the relevant parameters that satisfy the constraint of the SM-like Higgs boson measured mass in 3$\sigma$.

\section{125 GeV Higgs boson decays with lepton flavor violation in the B-LSSM\label{sec3}}
\begin{figure}
\setlength{\unitlength}{1mm}
\centering
\includegraphics[width=2.5in]{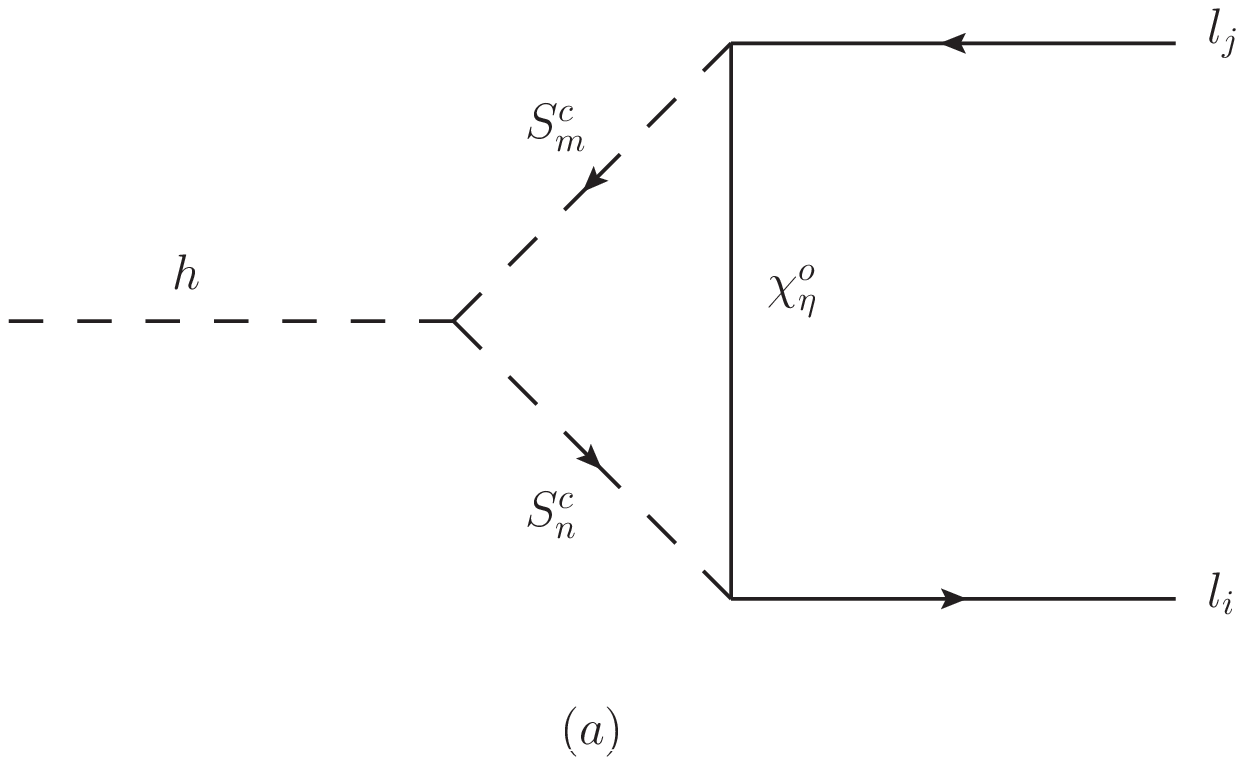}
\vspace{0cm}
\includegraphics[width=2.5in]{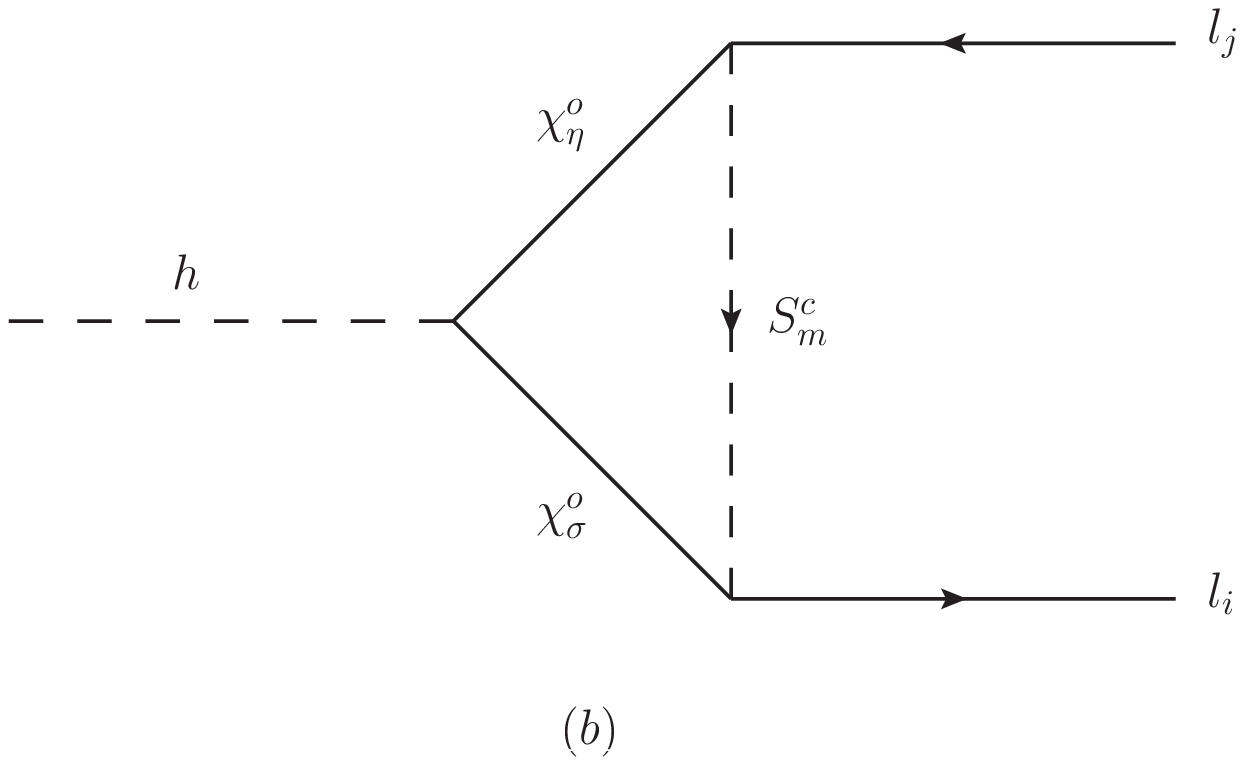}
\vspace{0cm}
\includegraphics[width=2.5in]{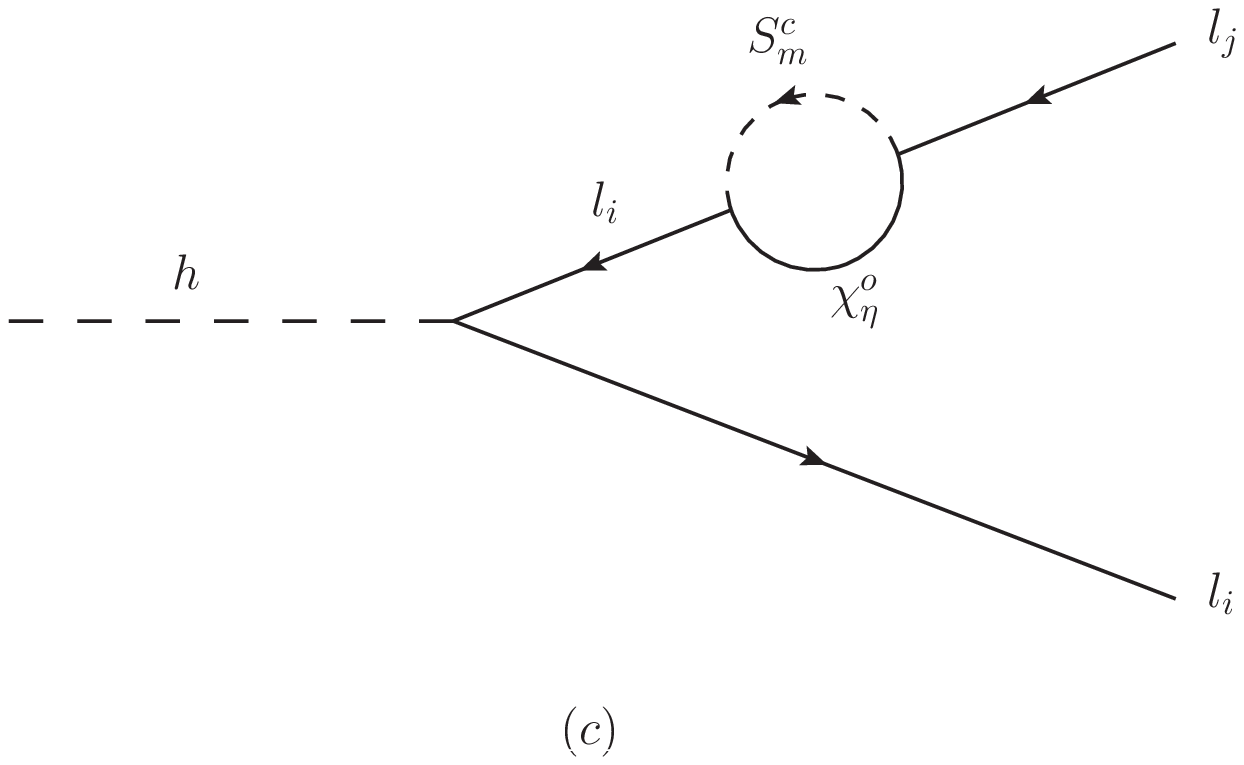}
\vspace{0cm}
\includegraphics[width=2.5in]{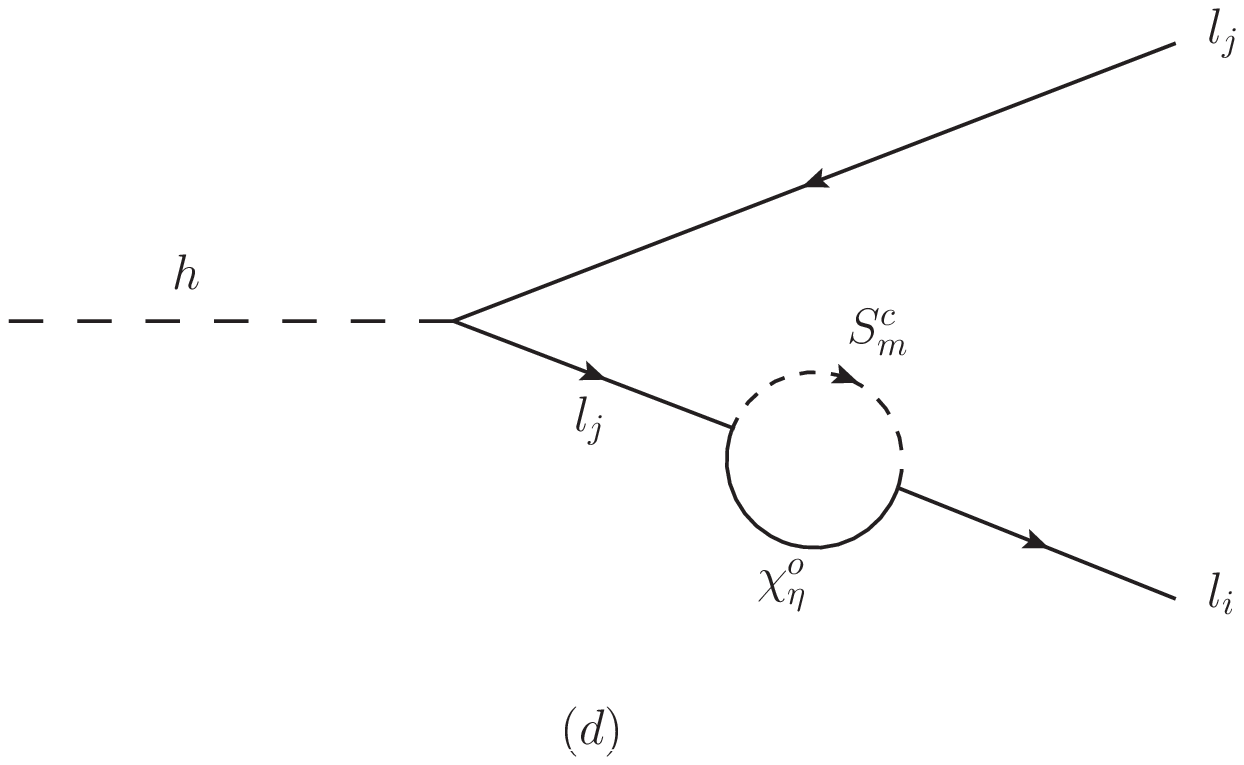}
\vspace{0cm}
\caption[]{{\label{1}} Feynman diagrams for the process $h\rightarrow \bar l_i l_j$ in the B-LSSM, where $S_{m,n}^c$ is charged scalar and $\chi_{\eta,\sigma}^o$ is neutral fermion. (a), (b) denote the contributions of vertex diagrams for $h\rightarrow \bar{l}_i l_j$ from loops. (c), (d) denote self-energy diagrams contributing to $h\rightarrow \bar{l}_i l_j$.}
\end{figure}

In this section, we analyze the decay width of 125 GeV Higgs boson decays with LFV $h\rightarrow \bar{l}_i l_j$ in the B-LSSM.
The effective amplitude of the Higgs boson decays $h\rightarrow \bar{l}_i l_j$ is generally written as
\begin{eqnarray}
&&\mathcal{M}=\bar{l}_i (F_L^{ij}P_L + F_R^{ij}P_R) l_j h,\\
&&F_{L,R}^{ij}=F_{L,R}^{(A)ij}+F_{L,R}^{(B)ij},
\end{eqnarray}
and $P_L=(1-\gamma_5)/2$, $P_R=(1+\gamma_5)/2$, $F_{L,R}^{(A)ij}$ is derived from the contributions of the vertex diagrams in Figs.~\ref{1}(a) and \ref{1}(b) and $F_{L,R}^{(B)ij}$ denotes the contributions from the self-energy diagrams in Figs.~\ref{1}(c) and \ref{1}(d).
\begin{eqnarray}
&&F_{L,R}^{(A)ij} =F_{L,R}^{(a)ij} +F_{L,R}^{(b)ij}, \\
&&F_{L,R}^{(B)ij} =F_{L,R}^{(c)ij} +F_{L,R}^{(d)ij}.
\end{eqnarray}
The formulas for neutral fermion loop contributions $F_{L,R}^{(a,b)ij}$ are as follows:
\begin{eqnarray}
&&\hspace{-0.75cm}F_L^{(a)ij}=  \frac{{m_{{\chi _\eta ^o}}}{C^{S^c}_{1mn}}}{{m_W^2}}
C_L^{S_n^c \chi _\eta ^o{{\bar l }_{i}}}
C_L^{S_m^{c*} {l_{j}}\bar \chi _\eta ^o}
{G_1}({x_{\chi _\eta ^o}},{x_{S_m^c}},{x_{S_n^c}}) ,\nonumber\\
&&\hspace{-0.75cm}F_L^{(b)ij}=  \frac{{m_{{\chi _\sigma^o }}}{m_{{\chi _\eta^o }}}}{{m_W^2}}
C_L^{{S_m^c}{\chi _\sigma^o }{{\bar l}_{i}}}C_L^{h{\chi _\eta^o }{{\bar \chi }_\sigma^o }}
C_L^{{S_m^{c*}}{l_{j}}{{\bar \chi }_\eta^o}}{G_1}({x_{{S_m^c}}},{x_{{\chi _\sigma^o }}},{x_{{\chi _\eta^o }}})\nonumber\\
&&\hspace{0.75cm} +\:  C_L^{{S_m^c}{\chi _\sigma^o }{{\bar l}_{i}}}C_R^{h{\chi _\eta^o }{{\bar \chi }_\sigma^o }}
C_L^{{S_m^{c*}}{l_{j}}{{\bar \chi }_\eta^o}}{G_2}({x_{{S_m^c}}},{x_{{\chi _\sigma^o }}},{x_{{\chi _\eta^o }}}) ,\nonumber\\
&&\hspace{-0.75cm}F_R^{(a,b)ij} = \left. {F_L^{(a,b)ij}} \right|{ _{L \leftrightarrow R}} ,
\end{eqnarray}
where $x_i=m_i^2/m_W^2$, $G_i$ is loop function and $C$ is coupling constant which can be found in the Appendices. The formulas for the self-energy diagrams contribution $F_{L,R}^{(B)ij}$ are as follows:
\begin{eqnarray}
&&\hspace{-0.75cm}F_{L}^{(c)ij} = \frac{C_L^{h l_i {\bar l}_i}}{m_{l_j}^2-m_{l_i}^2}
\Big\{ m_{l_j}^2 {\Sigma}_R (m_{l_j}^2) +  m_{l_j}^2 {\Sigma}_{Rs} (m_{l_j}^2)\nonumber\\
&&\hspace{0.7cm}
+\,  m_{l_i} [m_{l_j} {\Sigma}_L (m_{l_j}^2) + m_{l_j} {\Sigma}_{Ls} (m_{l_j}^2)] \Big\},\nonumber\\
&&\hspace{-0.75cm}F_{L}^{(d)ij} = \frac{C_L^{h l_j {\bar l}_j}}{m_{l_i}^2-m_{l_j}^2}
\Big\{ m_{l_i}^2 {\Sigma}_L (m_{l_i}^2) + m_{l_i}  m_{l_j} {\Sigma}_{Rs} (m_{l_i}^2) \nonumber\\
&&\hspace{0.7cm}
+\,  m_{l_j} [m_{l_i} {\Sigma}_R (m_{l_i}^2) + m_{l_j} {\Sigma}_{Ls} (m_{l_i}^2)] \Big\},\nonumber\\
&&\hspace{-0.75cm}F_{R}^{(c,d)ij} = \left. {F_{L}^{(c,d)ij}} \right|{ _{L \leftrightarrow R}}.
\end{eqnarray}
$\Sigma$ is loop contribution of self-energy diagrams,
\begin{eqnarray}
&&\hspace{-0.75cm}{\Sigma}_L (p^2) = -\frac{1}{16\pi^2}B_1(p^2,m_{\chi_\eta^o}^2,m_{S_m^c}^2)C_L^{S_m ^ c \chi _\eta ^o{{\bar l }_{i}}}C_R^{S_m ^{c*} {l_{j}}\bar \chi _\eta ^o},\nonumber\\
&&\hspace{-0.75cm}m_{l_j} {\Sigma}_{Ls} (p^2) = \frac{1}{16\pi^2} m_{\chi_\eta^o} B_0(p^2,m_{\chi_\eta^o}^2,m_{S_m^c}^2)C_L^{S_m ^ c \chi _\eta ^o{{\bar l }_{i}}}C_L^{S_m ^{c*} {l_{j}}\bar \chi _\eta ^o},\nonumber\\
&&\hspace{-0.75cm} {\Sigma}_{R} (p^2) = \left. {{\Sigma}_{L} (p^2)} \right|{ _{L \leftrightarrow R}},\nonumber\\
&&\hspace{-0.75cm} m_{l_j} {\Sigma}_{Rs} (p^2) = \left. {m_{l_j} {\Sigma}_{Ls} (p^2)} \right|{ _{L \leftrightarrow R}},
\end{eqnarray}
where $B_{0,1}(p^2,m_0^2,m_1^2)$ is the two-point function~\cite{ref-B-0,ref-B-1,ref-B-2,ref-B-3,ref-B-4,ref-B-5,ref-B-6}.

Then, we can obtain the decay width of $h\rightarrow \bar l_i l_j$~\cite{LFVHD-Br1,LFVHD-Br2},
\begin{eqnarray}
{\Gamma}(h\rightarrow \bar{l}_i l_j) \simeq \frac{m_h}{16\pi}\Big({\left| {F_L^{ij}} \right|^2} + {\left| {F_R^{ij}} \right|^2}\Big),
\end{eqnarray}
the decay width of $h\rightarrow l_i l_j$ is
\begin{eqnarray}
\Gamma(h\rightarrow l_i l_j)= {\Gamma}(h\rightarrow \bar{l}_i l_j)+{\Gamma}(h\rightarrow \bar{l}_j l_i ),
\end{eqnarray}
and the branching ratio of $h\rightarrow l_i l_j$ is
\begin{eqnarray}
{\rm{Br}}(h\rightarrow l_i l_j)= {\Gamma(h\rightarrow l_i l_j)}/{{\Gamma}_h}.
\end{eqnarray}
Here ${{\Gamma}}_h \approx {\Gamma}_h^{SM} \approx 4.1\times10^{-3}\:{\rm{GeV}}$~\cite{HCSWG}. ${\Gamma}_h$ denotes the total decay width of the 125 GeV Higgs boson in the B-LSSM. ${\Gamma}_h^{SM}$ is the predicted value of the 125 GeV Higgs boson total decay width in the SM. In the following numerical section, we choose the supersymmetric particles in the B-LSSM that are heavy and whose contributions to the decay width of the 125 GeV Higgs boson is weak. Hence, we choose ${\Gamma}_h$ is approximately equal to ${\Gamma}_h^{SM}$.

\section{Numerical analysis\label{sec4}}
First of all, in order to obtain reasonable numerical results, we need to study some sensitive parameters and important mass matrices. Then, to show the numerical results clearly, we will discuss the processes of $h\rightarrow e\mu$, $h\rightarrow e\tau$ and $h\rightarrow \mu\tau$ in three subsections.

The relevant SM input parameters are chosen as $m_W$\emph{=}$80.385{\rm GeV}$, $m_Z$\emph{=}$90.1876{\rm GeV}$, $\alpha_{em}(m_Z)=1/128.9$, $\alpha_s(m_Z)=0.118$. Considering that the updated experimental data  on searching $Z'$ indicate $M_{Z'}\geq 4.05 {\rm TeV}$ at 95\% C.L.~\cite{newZ}, we choose $M_{Z'}$\emph{=}$4.2 {\rm TeV}$ in the following. References~\cite{GCG,MAB} give an upper bound on the ratio between the $Z'$ mass and its gauge coupling at 99\% C.L. as $M_{Z'}/g_B\geq 6{\rm TeV}$, and then the scope of $g_B$ is $0<g_B<0.7$. LHC experimental data constrain $\tan\beta'<1.5$~\cite{48}. Considering the constraint of the experiments~\cite{PDGPA}, we take $M_{1}$\emph{=}$500{\rm GeV},\;M_{2}$\emph{=}$600{\rm GeV}$, $B_\mu'$\emph{=}$5\times10^5{\rm GeV}^2$, $m_{\tilde{q}}$\emph{=}$m_{\tilde{u}}$\emph{=}$diag(2, 2, 1.6){\rm TeV}$, and $T_u$\emph{=}$diag(1, 1, 1){\rm TeV}$, respectively.

We have to be clear that we are looking at 125 GeV Higgs boson decays with lepton flavor violation in the B-LSSM, so we need to consider the constraint of the SM-like Higgs boson mass. The remaining key parameters that affect the Higgs boson mass are $\tan\beta$, $\tan\beta'$, $g_B$, and $g_{YB}$. By constantly adjusting the parameters, the final numerical analysis strictly conforms to the constraint of the SM-like Higgs boson measured mass in 3$\sigma$. Finally, we know that the B-LSSM contains all the ingredients to induce nonzero neutrino mass, but since the neutrino mass is very small, it makes little contribution to the problem we study. So, here we choose that neutrino masses are zero in the numerical analysis. The other one is about the neutrino Yukawa sector. Although the B-LSSM contains LFV sources in the neutrino Yukawa sector, such as the $Y_\nu$ matrix, the neutrino oscillation causes $Y_\nu \sim \mathcal{O}(10^{-6})$, which contributes very little to the problem we study, so we approximately ignore the influence of the neutrino Yukawa sector in the numerical analysis.

After studying lepton flavor violating processes, we consider the off-diagonal terms for the soft breaking slepton mass matrices $m^2_{\bar L, \bar e}$ and the trilinear coupling matrix $T_e$, which are defined by~\cite{sl-mix,sl-mix1,sl-mix2,sl-mix3,sl-mix4,neu-zhang2}
\begin{eqnarray}
&&\hspace{-0.75cm}\quad\;\,{m^2_{\tilde L}} = \left( {\begin{array}{*{20}{c}}
   1 & \delta_{12}^{LL} & \delta_{13}^{LL}  \\
   \delta_{12}^{LL} & 1 & \delta_{23}^{LL}  \\
   \delta_{13}^{LL} & \delta_{23}^{LL} & 1  \\
\end{array}} \right){m_L^2},\\
&&\hspace{-0.75cm}\quad\:{m_{\tilde e^c}^2} = \left( {\begin{array}{*{20}{c}}
   1 & \delta_{12}^{RR} & \delta_{13}^{RR}  \\
   \delta_{12}^{RR} & 1 & \delta_{23}^{RR}  \\
   \delta_{13}^{RR} & \delta_{23}^{RR} & 1  \\
\end{array}} \right){m_E^2},\\
&&T_e=
\left(\begin{array}{ccc} 1& \delta_{12}^{LR} &\delta_{13}^{LR}
\\ \delta_{12}^{LR} &1 &\delta_{23}^{LR}
\\ \delta_{13}^{LR} &\delta_{23}^{LR}  &1\end{array}\right){A_e}.
\end{eqnarray}
Here, the definition of the slepton flavor mixing is more detailed than that of in our previous paper~\cite{50} which studied the lepton flavor violating decays $l_j^-\rightarrow l_i^-\gamma$ and $l_j^-\rightarrow l_i^-l_i^-l_i^+$ in the B-LSSM. In our previous paper~\cite{50}, we just considered the off-diagonal terms for the trilinear coupling matrix $T_e$.
In the subsequent numerical analysis, we will show that the branching ratios of $h\rightarrow e\mu$, $h\rightarrow e\tau$, and $h\rightarrow \mu\tau$ in the B-LSSM depend on the slepton mixing parameters $\delta_{12}^{XX}$, $\delta_{13}^{XX}$, and $\delta_{23}^{XX}~(X=L,R)$, respectively, constrained by the present experimental limits on the branching ratio of $l_j^-\rightarrow l_i^-\gamma$.

We also impose a constraint on the NP contribution to the muon anomalous magnetic dipole moment (MDM), $a_\mu=(g-2)_\mu$, in the B-LSSM~\cite{50,MDMMDM}. The new experimental average for the difference between the experimental measurement and SM theoretical prediction of $a_\mu$ now is given by~\cite{MDM-exp1}
\begin{eqnarray}
\Delta a_\mu =a_\mu^{{\rm{exp}}} -a_\mu^{{\rm{SM}}} = (25.1\pm5.9)\times 10^{-10}.
\label{MDM-exp}
\end{eqnarray}
Therefore, the NP contribution to the muon anomalous MDM in the B-LSSM,  $\Delta a^{NP}_\mu$, should be constrained as $13.3\times 10^{-10} \leq \Delta a^{NP}_\mu \leq 36.9\times 10^{-10}$, where we consider 2$\sigma$ experimental error.

We all know that LFV processes are flavor dependent, just as LFV rate for $e\rightarrow \mu$ transitions depends on the slepton mixing parameters $\delta_{12}^{XX}~(X=L,R)$, which can be confirmed by Fig.~\ref{2}. In the following, we choose  $\tan\beta=11$, $\tan\beta'=1.3$, $g_B=0.5$, $g_{YB}=-0.4$, $m_L=m_E=1$ TeV, and $A_e=0.5$ TeV, unless it is a variable in a graph. Note that, when we look at the effect of $\delta_{12}^{XX}~(X=L,R)$ on ${\rm{Br}}(h\rightarrow e\mu)$, we keep the other two slepton mixing parameters $\delta_{12}^{XX}~(X=L,R)=0$. The slepton mixing parameters $\delta_{13}^{XX}~(X=L,R)$ and $\delta_{23}^{XX}~(X=L,R)$ hardly influence the LFV rates for $e\rightarrow \mu$ transitions, so we take $\delta_{13}^{XX}=0$ and $\delta_{23}^{XX}=0$ $(X=L, R)$ in subsection IV A. The same goes for the other two processes of LFV rates for $ e\rightarrow \tau$ and $\mu\rightarrow \tau$ transitions in subsections IV B and IV C.

\subsection{125 GeV Higgs boson decay with lepton flavor violation  $h\rightarrow e\mu $ }
\begin{figure}
\setlength{\unitlength}{1mm}
\centering
\includegraphics[width=2.4in]{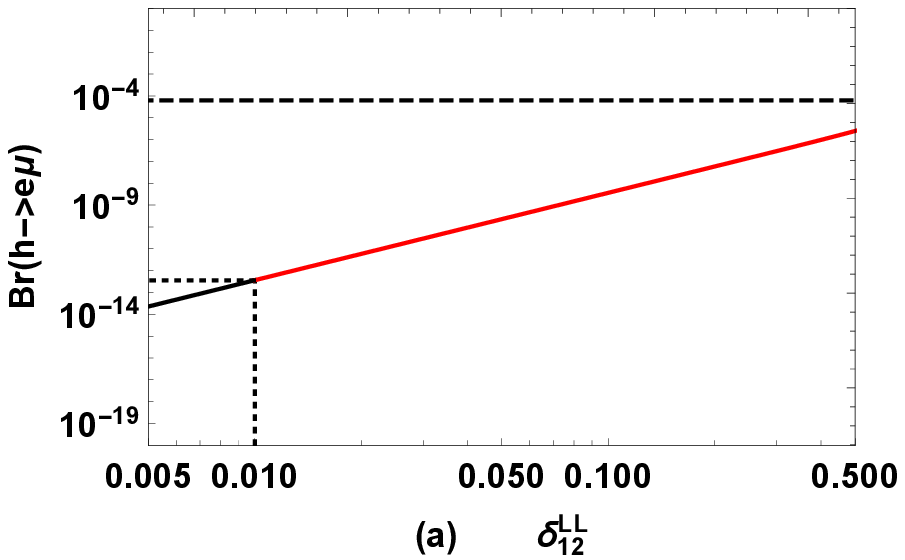}
\vspace{0cm}
\includegraphics[width=2.4in]{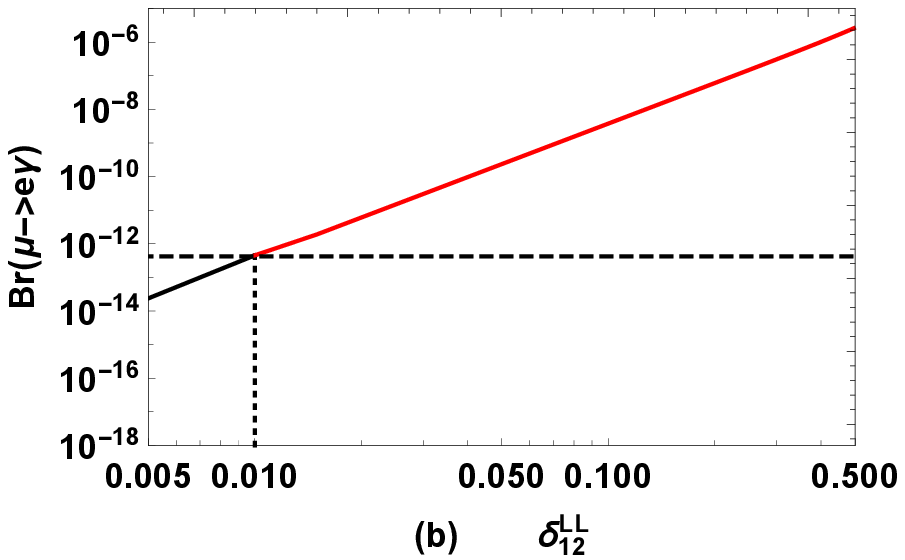}
\vspace{0cm}
\includegraphics[width=2.4in]{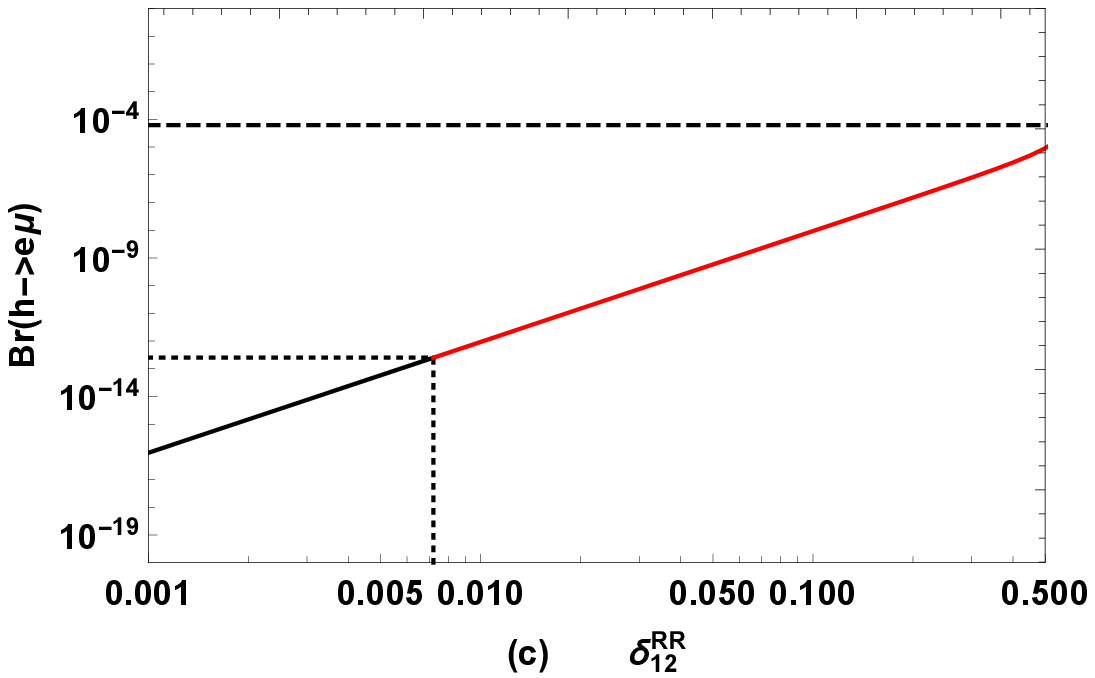}
\vspace{0cm}
\includegraphics[width=2.4in]{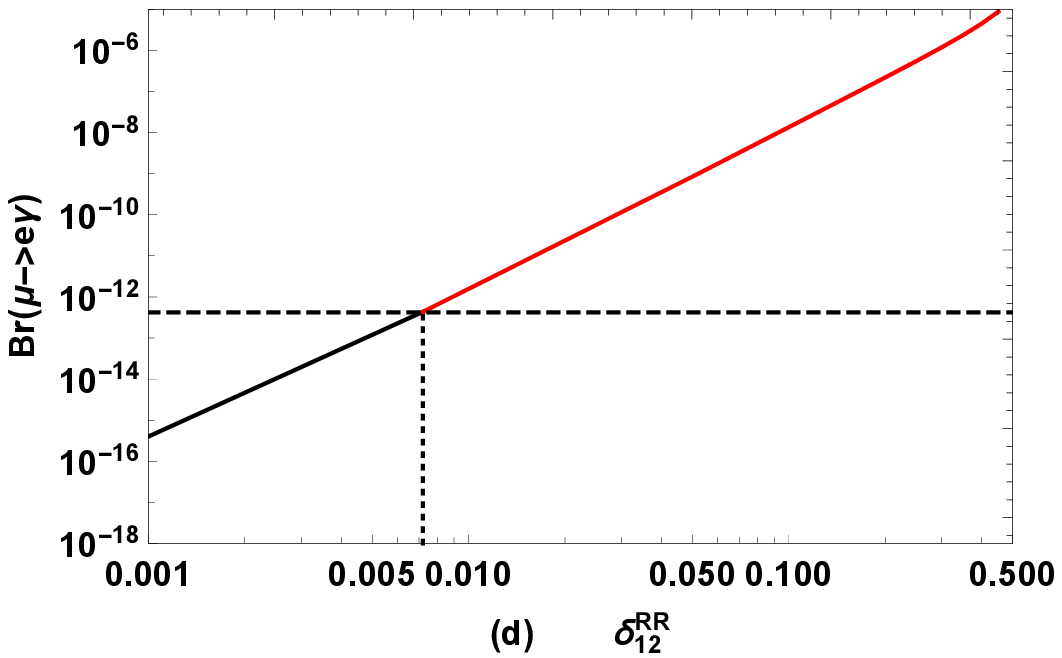}
\vspace{0cm}
\includegraphics[width=2.4in]{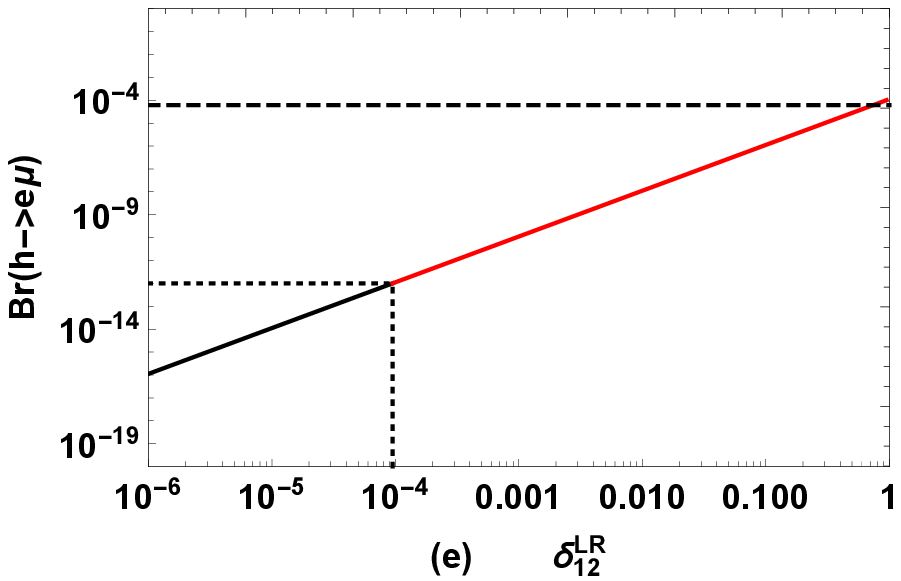}
\vspace{0cm}
\includegraphics[width=2.4in]{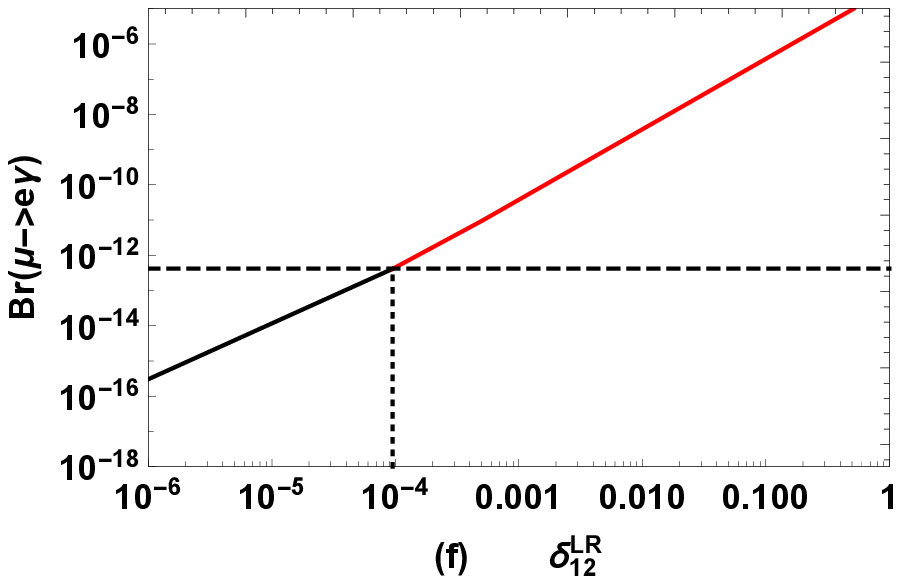}
\vspace{0cm}
\caption[]{{\label{2}} ${\rm{Br}}(h\rightarrow e\mu)$  versus slepton flavor mixing parameters $\delta_{12}^{LL}$ (a), $\delta_{12}^{RR}$ (c), and $\delta_{12}^{LR}$ (e), where the dashed line stands for the upper limit on ${\rm{Br}}(h\rightarrow e\mu )$ at 95\% C.L. as shown in Eq.(\ref{a1}). ${\rm{Br}}(\mu \rightarrow e\gamma)$ versus slepton mixing parameters $\delta_{12}^{LL}$ (b), $\delta_{12}^{RR}$ (d), and $\delta_{12}^{LR}$ (f), where the dashed line denotes the present limit of ${\rm{Br}}(\mu\rightarrow e\gamma)$ at 90\% C.L. as shown in Eq.(\ref{a4}). Here, the red solid line is ruled out by the present limit of ${\rm{Br}}(\mu \rightarrow e\gamma)$, and the black solid line is consistent with the present limit of ${\rm{Br}}(\mu \rightarrow e\gamma)$.}
\end{figure}

In this subsection, we mainly analyze 125 GeV Higgs boson decays with LFV $h\rightarrow e\mu$ in the B-LSSM. First, we plot the influence of slepton flavor mixing parameters $\delta^{XX}_{12}$ $(X=L, R)$ on the branching ratios of $h\rightarrow e\mu$ and $ \mu\rightarrow e\gamma$ in Fig.~\ref{2}. In the figure, the latest experimental upper limits of the branching ratios of $h\rightarrow e\mu$ and $ \mu\rightarrow e\gamma$ are displayed as the dashed lines, the red solid line is ruled out by the present limit of ${\rm{Br}}(\mu \rightarrow e\gamma)$, and the black solid line is consistent with the present limit of ${\rm{Br}}(\mu \rightarrow e\gamma)$.

In Fig.~\ref{2}, we can clearly see that when the slepton mixing parameters close $\delta_{12}^{XX} $(X=L, R)$=0$, ${\rm{Br}}(h\rightarrow e\mu)$ can be approximately  $\mathcal{O}(10^{-16})$, which is too small to be detected experimentally. However, with the slepton mixing parameters, $\delta_{12}^{XX}$ $(X=L, R)$ increases, both ${\rm{Br}}(h\rightarrow e\mu)$ and ${\rm{Br}}(\mu\rightarrow e\gamma)$ will grow rapidly, and ${\rm{Br}}(\mu\rightarrow e\gamma)$ will soon exceed the experimental limit. Although ${\rm{Br}}(h\rightarrow e\mu)$ does not exceed the experimental limit, it will also be very close to the experimental limit. Especially in Fig.~\ref{2}(e), ${\rm{Br}}(h\rightarrow e\mu)$ will reach the experimental upper limit with the increase of  $\delta_{12}^{LR}$. As we can see from Fig.~\ref{2}, the influence of slepton flavor mixing parameters on ${\rm{Br}}(h\rightarrow e\mu)$ and ${\rm{Br}}(\mu\rightarrow e\gamma)$ is huge. The reason is that LFV processes are flavor dependent, and LFV rate for $e\rightarrow \mu$ transitions depends on the slepton mixing parameters $\delta_{12}^{XX}~(X=L,R)$. When we consider the constraint of ${\rm{Br}}(\mu\rightarrow e\gamma)$ to ${\rm{Br}}(h\rightarrow e\mu)$, ${\rm{Br}}(h\rightarrow e\mu)$ can be up to $\mathcal{O}(10^{-12})$. Therefore, it can be seen that the limit of ${\rm{Br}}(\mu\rightarrow e\gamma)$ to ${\rm{Br}}(h\rightarrow e\mu)$ is very strict, which makes ${\rm{Br}}(h\rightarrow e\mu)$ difficult to reach the experimental upper limit.

\begin{figure}
\setlength{\unitlength}{2mm}
\centering
\includegraphics[width=2.4in]{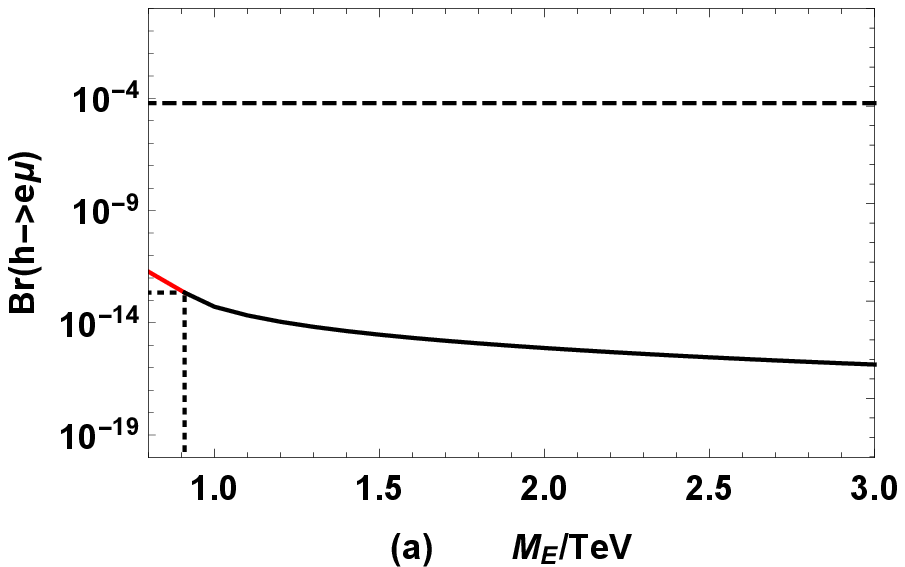}
\vspace{0cm}
\includegraphics[width=2.4in]{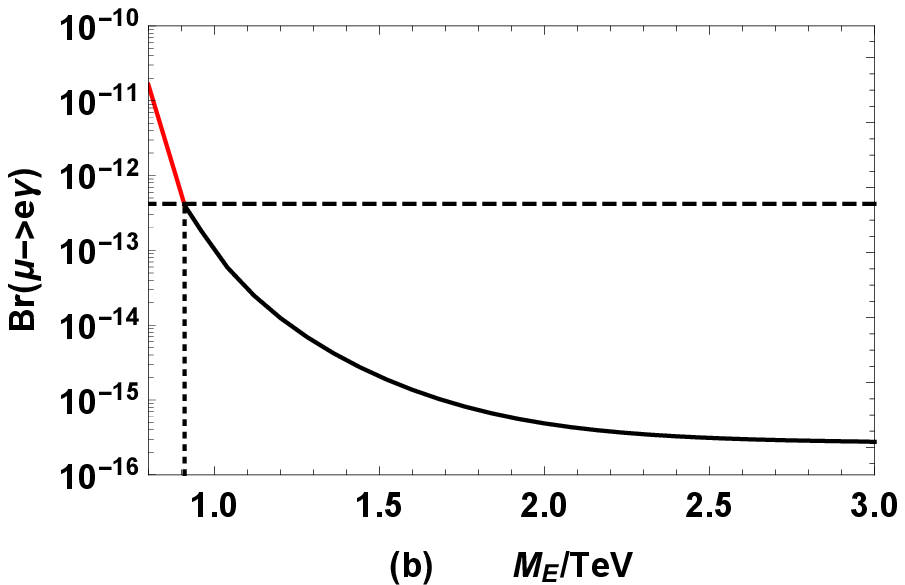}
\vspace{0cm}
\includegraphics[width=2.4in]{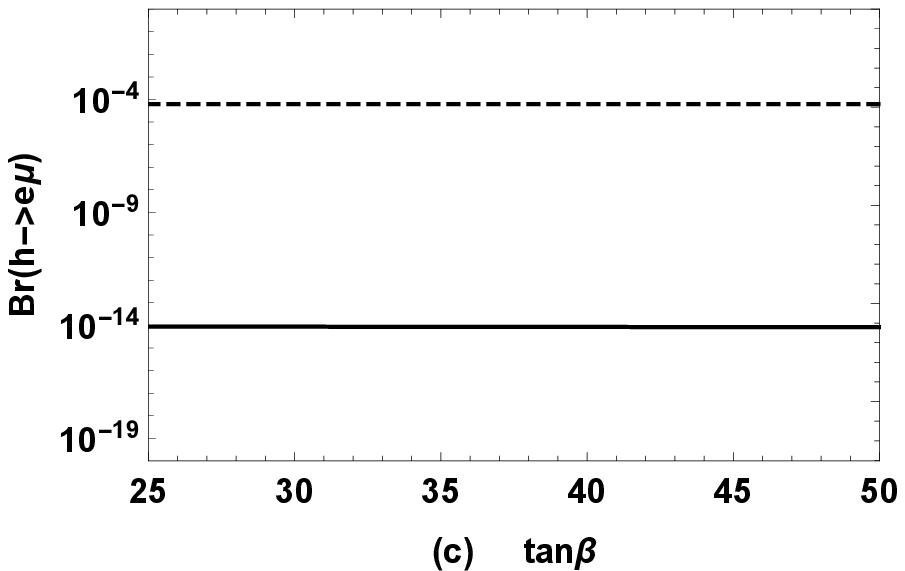}
\vspace{0cm}
\includegraphics[width=2.4in]{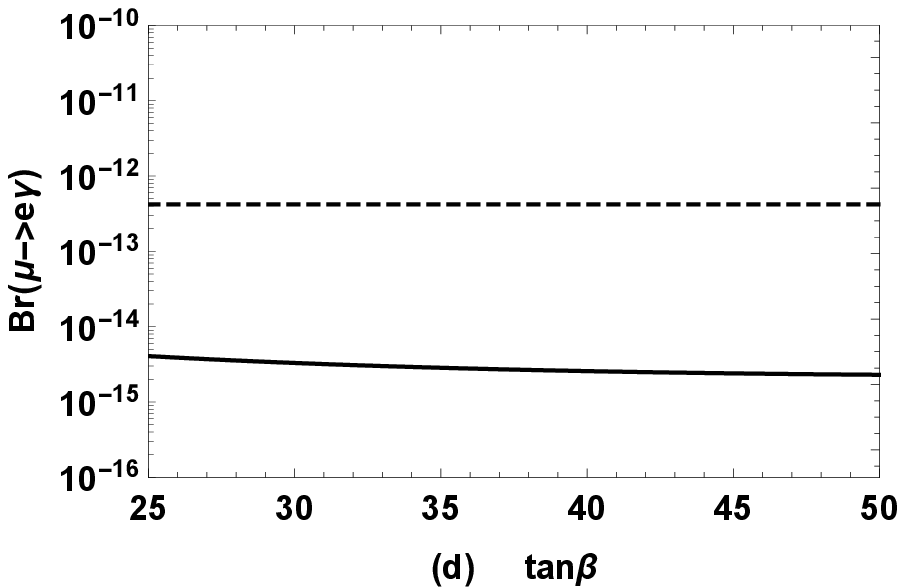}
\vspace{0cm}
\includegraphics[width=2.4in]{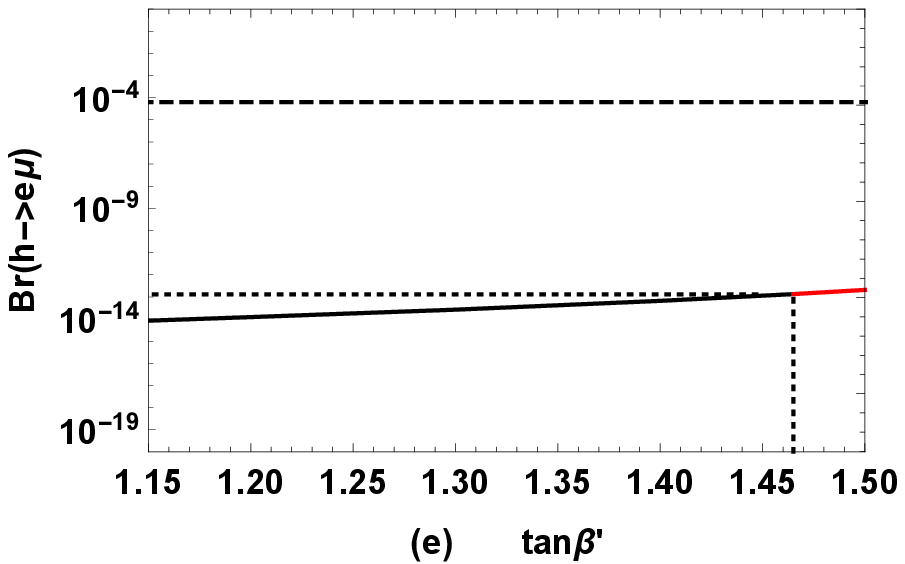}
\vspace{0cm}
\includegraphics[width=2.4in]{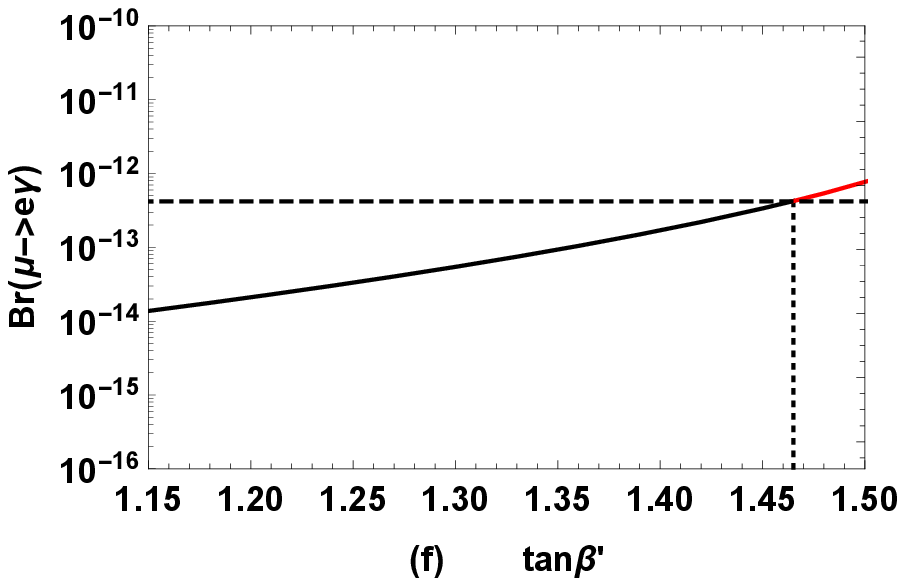}
\vspace{0cm}
\includegraphics[width=2.4in]{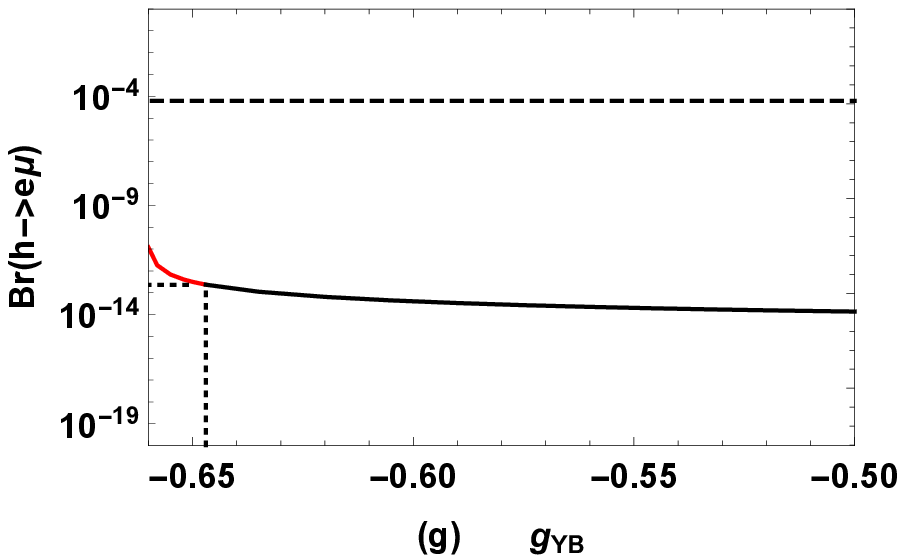}
\vspace{0cm}
\includegraphics[width=2.4in]{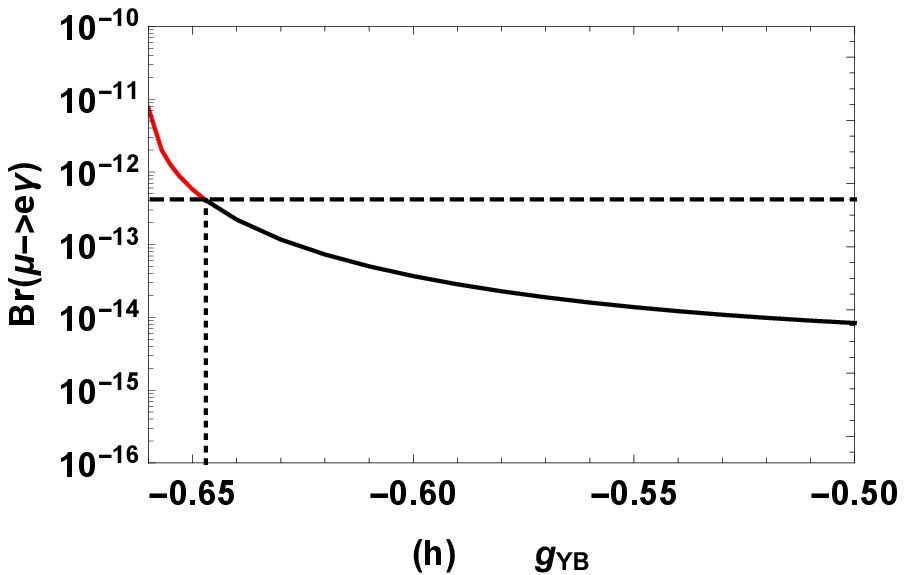}
\vspace{0cm}
\caption[]{{\label{3}} ${\rm{Br}}(h\rightarrow e\mu)$ versus other basic parameters $M_E$ (a), $\tan\beta$ (c), $\tan\beta'$ (e), and $g_{YB}$ (g), where the dashed line stands for the upper limit on ${\rm{Br}}(h\rightarrow e\mu )$ at 95\% C.L. as shown in Eq.(\ref{a1}). ${\rm{Br}}(\mu \rightarrow e\gamma)$ versus other basic parameters $M_E$ (b), $\tan\beta$ (d), $\tan\beta'$ (f), and $g_{YB}$ (h), where the dashed line denotes the present limit of ${\rm{Br}}(\mu\rightarrow e\gamma)$ at 90\% C.L. as shown in Eq.(\ref{a4}). The red solid line is ruled out by the present limit of ${\rm{Br}}(\mu \rightarrow e\gamma)$, and the black solid line is consistent with the present limit of ${\rm{Br}}(\mu \rightarrow e\gamma)$.}
\end{figure}

In addition, we study the influence of other basic parameters on ${\rm{Br}}(h\rightarrow e\mu)$ and ${\rm{Br}}(\mu\rightarrow e\gamma)$. We first set appropriate numerical values for slepton flavor mixing parameters, such as $\delta_{12}^{LL}=6\times 10^{-3}$, $\delta_{12}^{RR}=3\times 10^{-3}$, and $\delta_{12}^{LR}=1\times10^{-5}$. We also keep neutral fermion masses $m_{\chi_\eta^o}>200{\rm GeV}$ $(\eta=1,\cdots,7)$, the chargino masses $m_{\chi_{\alpha, \beta}}>200{\rm GeV}$ and the scalar masses $m_{S_{m, n}^{c}}>500{\rm GeV}$ $(m,n=1,\cdots,6)$ to avoid the range ruled out by the experiments. Then we research the influence of the basic parameters $M_E$, $\tan\beta$, $\tan\beta'$, and $g_{YB}$ on ${\rm{Br}}(h\rightarrow e\mu)$ and ${\rm{Br}}( \mu\rightarrow e\gamma)$, respectively, which can be intuitively seen in Fig.~\ref{3}.

The dashed line in Fig.~\ref{3} still represents the upper limit of the experiment. It is obvious that ${\rm{Br}}(h\rightarrow e\mu)$ and ${\rm{Br}}(\mu\rightarrow e\gamma)$ decrease with the increasing of $M_E$, the mass of sleptons increases as $M_E$ increases, which indicates that heavy sleptons play a suppressive role in the rates of LFV processes. Figures.~\ref{3} (a) and \ref{3} (b) show that the present experimental limit bounds of ${\rm{Br}}(\mu\rightarrow e\gamma)$ constrain $M_E\gtrsim0.9{\rm TeV}$. Figures.~\ref{3} (c) and \ref{3} (d) show that the LFV rates decrease with the increase of $\tan\beta$. The effect of $\tan\beta$ on ${\rm{Br}}(h\rightarrow e\mu)$ and ${\rm{Br}}(\mu\rightarrow e\gamma)$ is relatively small, which fails to make the two processes reach the experimental upper limit. $\tan\beta$ affects the numerical results mainly through the new mass matrix of Higgs bosons, neutralino and sleptons.

Compared to the MSSM, $\tan\beta'$ and $g_{YB}$ are new parameters in the B-LSSM. We plot ${\rm{Br}}(h\rightarrow e\mu)$ and ${\rm{Br}}(\mu\rightarrow e\gamma)$ versus $\tan\beta'$ and $g_{YB}$ in Figs.~\ref{3} (e)-\ref{3}(h), respectively. Figuers.~\ref{3} (e) and \ref{3} (f) show that LFV rates increase with the increasing of $\tan\beta'$. ${\rm{Br}}(\mu\rightarrow e\gamma)$ can reach the experimental upper limit but ${\rm{Br}}(h\rightarrow e\mu)$ cannot do that. As can be seen from the figure, the influence of $\tan\beta'$ on ${\rm{Br}}(\mu\rightarrow e\gamma)$ is more obvious than that of ${\rm{Br}}(h\rightarrow e\mu)$. The main reason is that the two decay processes contain different coupling vertices; for example, the coupling vertices of the Higgs boson and two sleptons are not the same as the coupling vertices of the photon and two sleptons, and $\tan\beta'$ affects the numerical results mainly through the new mass matrix of sleptons. $g_{YB}$ is also a new parameter in the B-LSSM; it can be seen from Figs.~\ref{3} (g) and \ref{3} (h) that ${\rm{Br}}(h\rightarrow e\mu)$ and ${\rm{Br}}(\mu\rightarrow e\gamma)$ decrease with the increase of $g_{YB}$. When $g_{YB}$ is small, ${\rm{Br}}(\mu\rightarrow e\gamma)$ can reach the experimental upper limit, but ${\rm{Br}}(h\rightarrow e\mu)$ cannot reach the experimental upper limit. $g_{YB}$ affect the numerical results through the new mass matrix of sleptons, Higgs bosons, and neutralino, which can make contributions to these LFV processes.

\begin{table*}
\begin{tabular*}{\textwidth}{@{\extracolsep{\fill}}lllll@{}}
\hline
Parameters&Min&Max&Step\\
\hline
$\tan\beta$&25&45&1&\\
$g_{YB}$&-0.7&-0.1&0.1&\\
$g_{B}$&0.1&0.7&0.1&\\
$\tan\beta'$&1&1.5&0.1&\\
\hline
$\delta_{12}^{XX}$&$10^{-5}$&$10^{-3}$&$5\times10^{-5}$\\
$\delta_{13}^{XX}$&0.05&0.90&0.05\\
$\delta_{23}^{XX}$&0.05&0.90&0.05\\
\hline
\end{tabular*}
\caption{Scanning parameters for Figs.~\ref{8}, ~\ref{9}, and ~\ref{10}}
\label{tab2}
\end{table*}

\begin{figure}
\setlength{\unitlength}{0mm}
\centering
\includegraphics[width=2.6in]{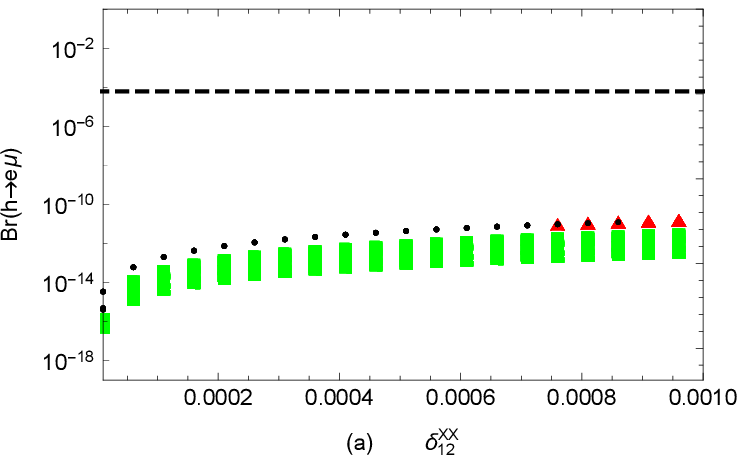}
\vspace{0cm}
\includegraphics[width=2.6in]{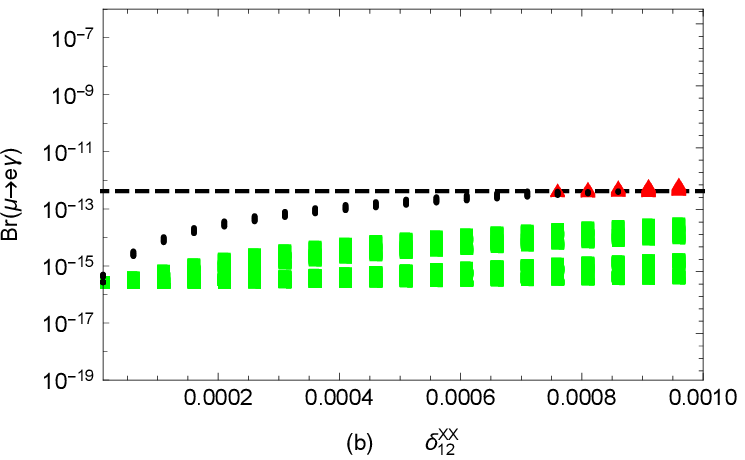}
\vspace{0cm}
\includegraphics[width=2.6in]{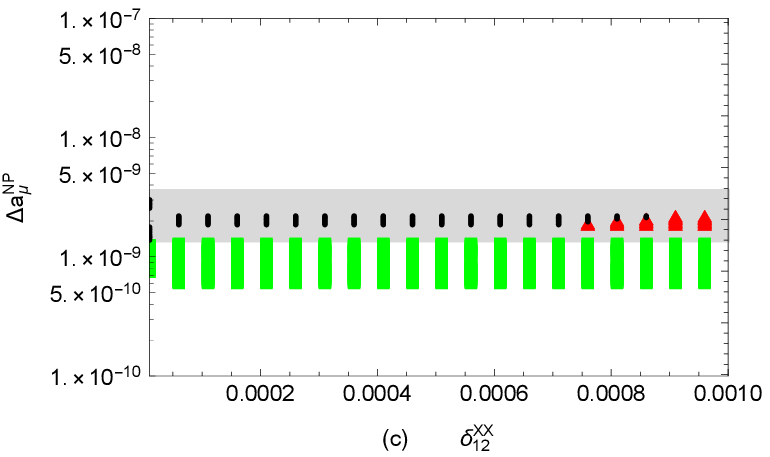}
\vspace{0cm}
\caption[]{{\label{8}} ${\rm{Br}}(h\rightarrow e\mu)$ versus slepton flavor mixing parameters $\delta_{12}^{XX}$ (a), where the dashed line stands for the upper limit on ${\rm{Br}}(h\rightarrow e\mu )$ at 95\% C.L. as shown in Eq.(\ref{a1}). ${\rm{Br}}(\mu \rightarrow e\gamma)$ versus slepton flavor mixing parameters $\delta_{12}^{XX}$ (b), where the dashed line denotes the present limit of ${\rm{Br}}(\mu\rightarrow e\gamma)$ at 90\% C.L. as shown in Eq.(\ref{a4}). $\Delta a^{NP}_\mu$ versus slepton flavor mixing parameters $\delta_{12}^{XX}$ (c), where the gray area denotes the $\Delta a_\mu$ at $2\sigma$ given in Eq.(\ref{MDM-exp}). Here, the red triangles are excluded by the present limit of ${\rm{Br}}(\mu \rightarrow e\gamma)$, the green squares are eliminated by $\Delta a_\mu$ at $2\sigma$, and the black circles simultaneously conform to the present limit of ${\rm{Br}}(\mu \rightarrow e\gamma)$ and the $\Delta a_\mu$ at $2\sigma$.}
\end{figure}

In order to clearly see the constraints of ${\rm{Br}}(\mu\rightarrow e\gamma)$ and $\Delta a_\mu$ on ${\rm{Br}}(h\rightarrow e\mu)$, we scan the parameter space shown in Table~\ref{tab2}. We set $\delta_{12}^{LL}=\delta_{12}^{RR}=\delta_{12}^{LR}=\delta_{12}^{XX}$, $A_e=0.5$ TeV, $m_E=0.5$ TeV. Under the condition that the SM-like Higgs boson mass $m_h$\emph{=}$125.10\pm 0.14\: {\rm{GeV}}$ in 3$\sigma$, neutral fermion masses $m_{\chi_\eta^o}>200{\rm GeV}$ $(\eta=1,\cdots,7)$, chargino masses $m_{\chi_{\alpha, \beta}}>200{\rm GeV}$ and the scalar masses $m_{S_{m, n}^{c}}>500{\rm GeV}$ $(m,n=1,\cdots,6)$ are satisfied, we obtain the relation of ${\rm{Br}}(h\rightarrow e\mu)$, ${\rm{Br}}(\mu\rightarrow e\gamma)$, $\Delta a^{NP}_\mu$ versus $\delta_{12}^{XX}$, respectively, as shown in Fig.~\ref{8}. The same analysis was also carried out in subsections IV B and TV C.

In Fig.~\ref{8}, the dashed line represents the upper limit of the experiment, the red triangles are excluded by the present limit of ${\rm{Br}}(\mu \rightarrow e\gamma)$, the green squares are eliminated by the $\Delta a_\mu$ at $2\sigma$, and the black circles simultaneously conform to the present limit of ${\rm{Br}}(\mu \rightarrow e\gamma)$ and the $\Delta a_\mu$ at $2\sigma$. It can be intuitively seen that ${\rm{Br}}(\mu \rightarrow e\gamma)$ and $\Delta a_\mu$ have a strict limitation on ${\rm{Br}}(h\rightarrow e\mu)$. Under the constraints of ${\rm{Br}}(\mu \rightarrow e\gamma)$ and $\Delta a_\mu$, ${\rm{Br}}(h\rightarrow e\mu)$ can reach $\mathcal{O}(10^{-11})$. That means that ${\rm{Br}}(h\rightarrow e\mu)$ now is very hard to get to the upper limit of the experiment.

\subsection{125 GeV Higgs boson decays with lepton flavor violation  $h\rightarrow e\tau $ }

\begin{figure}
\setlength{\unitlength}{1mm}
\centering
\includegraphics[width=2.4in]{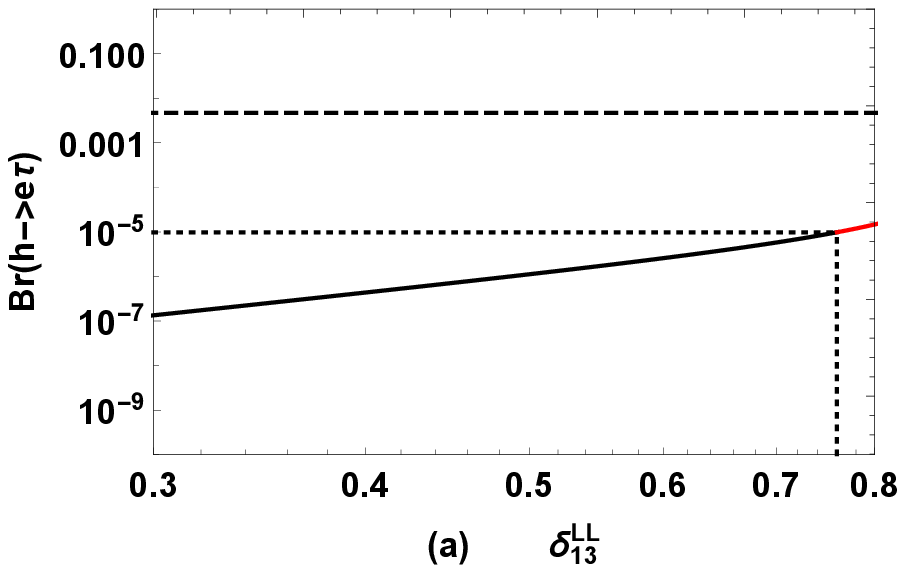}
\vspace{0cm}
\includegraphics[width=2.4in]{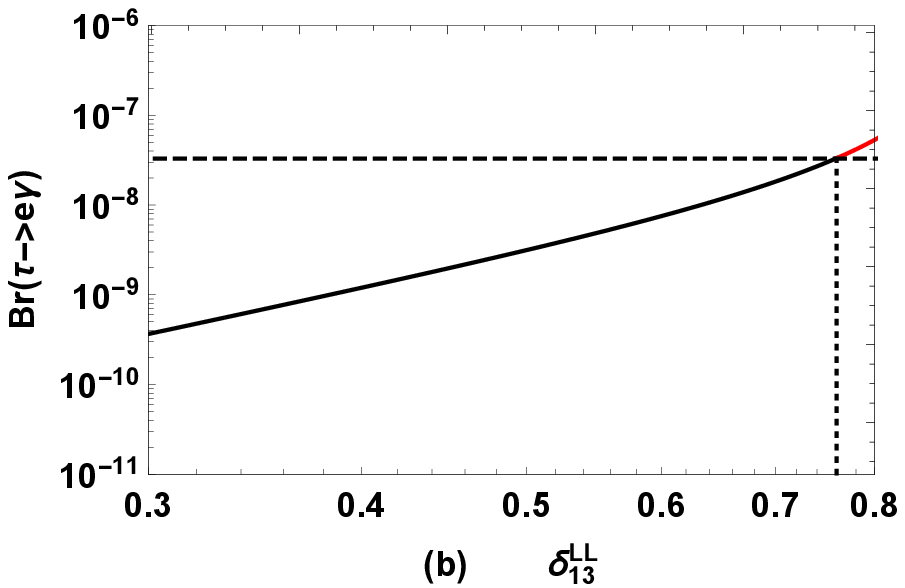}
\vspace{0cm}
\includegraphics[width=2.4in]{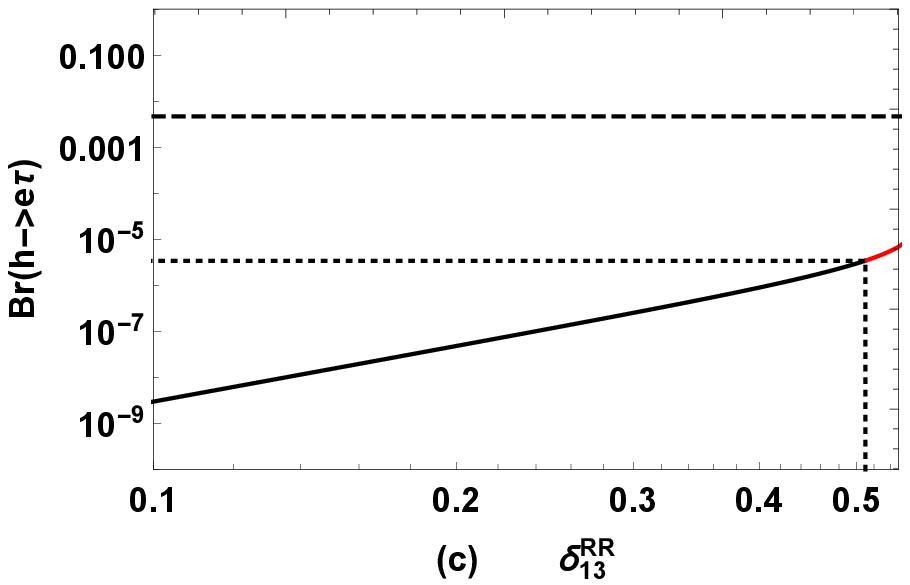}
\vspace{0cm}
\includegraphics[width=2.4in]{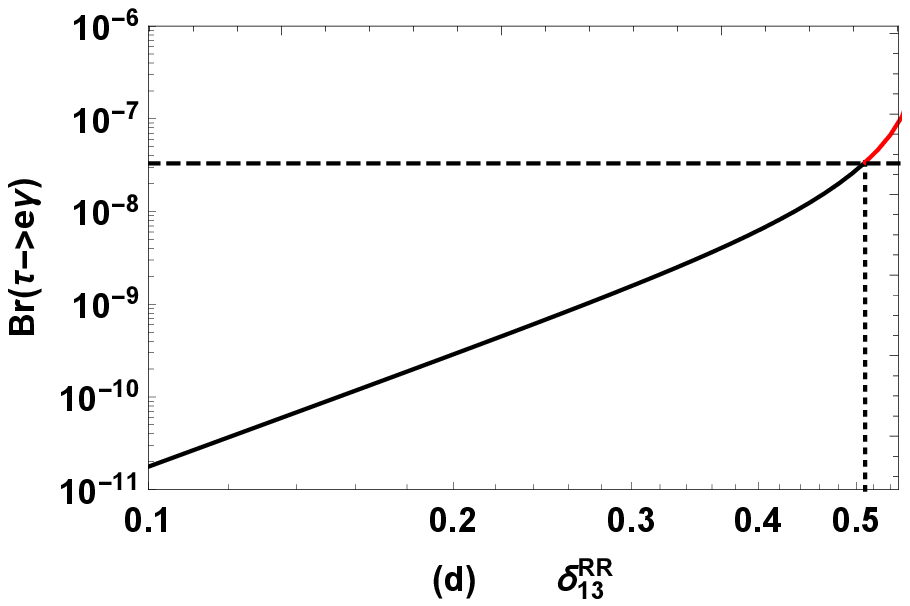}
\vspace{0cm}
\includegraphics[width=2.4in]{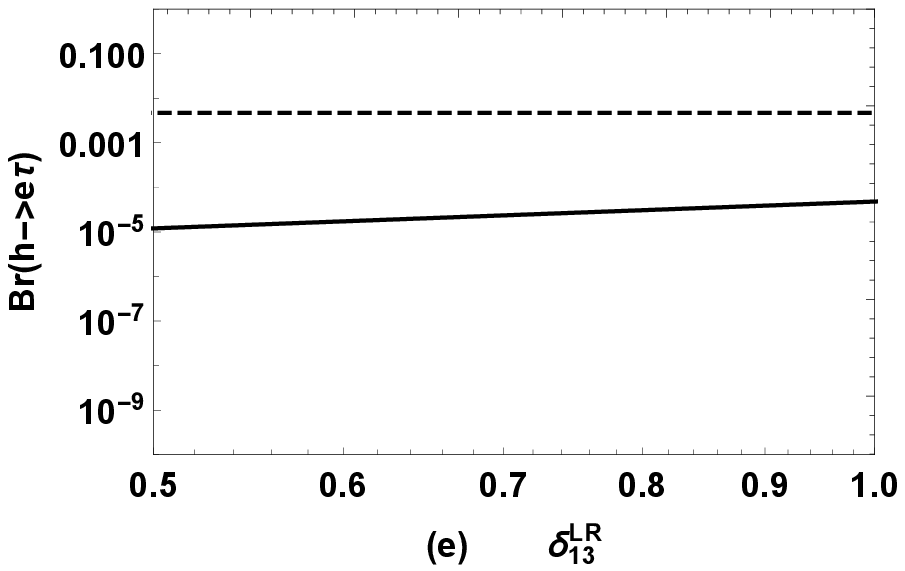}
\vspace{0cm}
\includegraphics[width=2.4in]{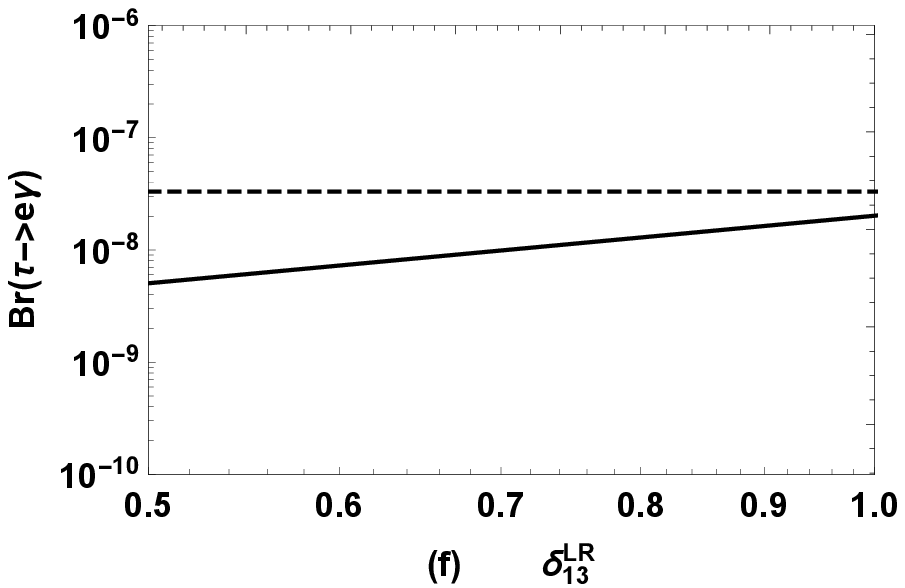}
\vspace{0cm}
\caption[]{{\label{4}} ${\rm{Br}}(h\rightarrow e\tau)$  versus slepton flavor mixing parameters $\delta_{13}^{LL}$ (a), $\delta_{13}^{RR}$ (c), and $\delta_{13}^{LR}$ (e), where the dashed line stands for the upper limit on ${\rm{Br}}(h\rightarrow e\tau )$ at 95\% C.L. as shown in Eq.(\ref{a2}). ${\rm{Br}}(\tau \rightarrow e\gamma)$ versus slepton mixing parameters $\delta_{13}^{LL}$ (b), $\delta_{13}^{RR}$ (d), and $\delta_{13}^{LR}$ (f), where the dashed line denotes the present limit of ${\rm{Br}}(\tau \rightarrow e\gamma)$ at 90\% C.L. as shown in Eq.(\ref{a5}). The red solid line is ruled out by the present limit of ${\rm{Br}}(\tau \rightarrow e\gamma)$, and the black solid line is consistent with the present limit of ${\rm{Br}}(\tau \rightarrow e\gamma)$.}
\end{figure}

In this section, we analyze the 125 GeV Higgs boson decays with LFV $h\rightarrow e\tau$ in the B-LSSM. In Fig.~\ref{4}, we picture ${\rm{Br}}(h\rightarrow e\tau) $ and ${\rm{Br}}(\tau \rightarrow e\gamma)$ varying with the slepton flavor mixing parameter $\delta_{13}^{XX}$ $(X=L, R) $, where the dashed lines denote the latest experimental upper limits of ${\rm{Br}}(h\rightarrow e\tau) $ and ${\rm{Br}}(\tau \rightarrow e\gamma)$.

\begin{figure}
\setlength{\unitlength}{2mm}
\centering
\includegraphics[width=2.4in]{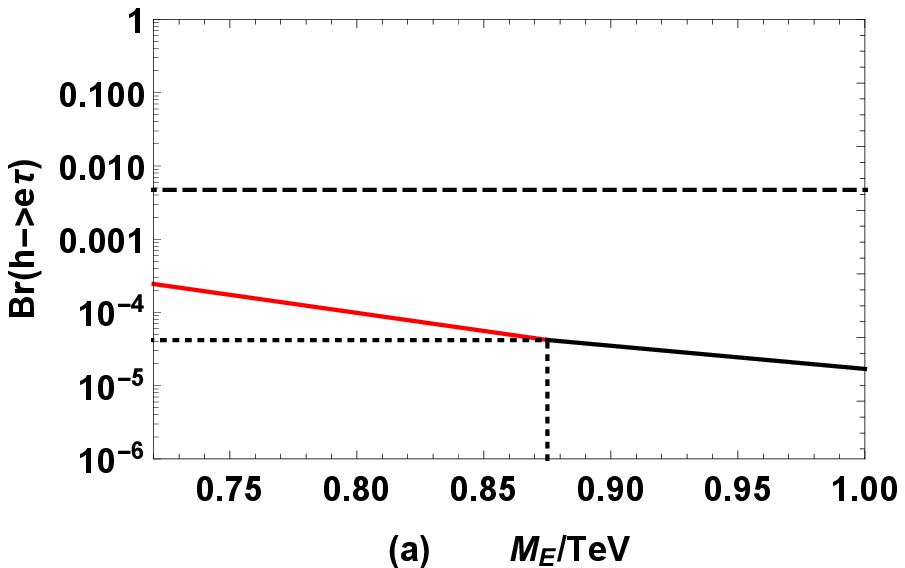}
\vspace{0cm}
\includegraphics[width=2.4in]{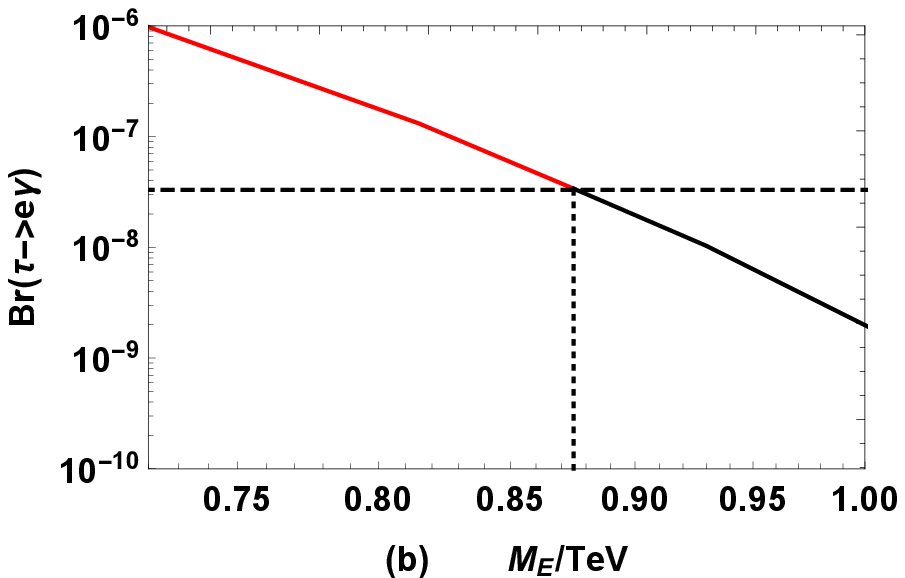}
\vspace{0cm}
\includegraphics[width=2.4in]{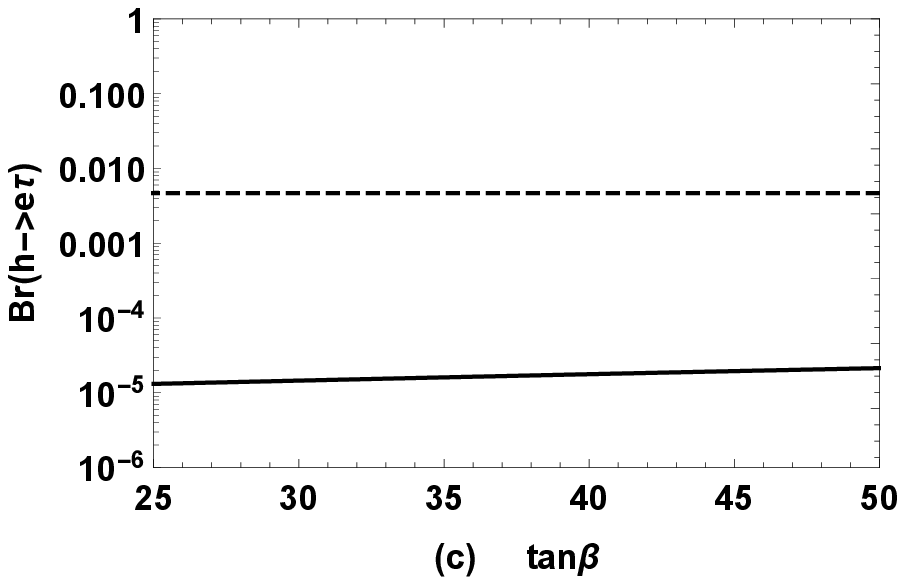}
\vspace{0cm}
\includegraphics[width=2.4in]{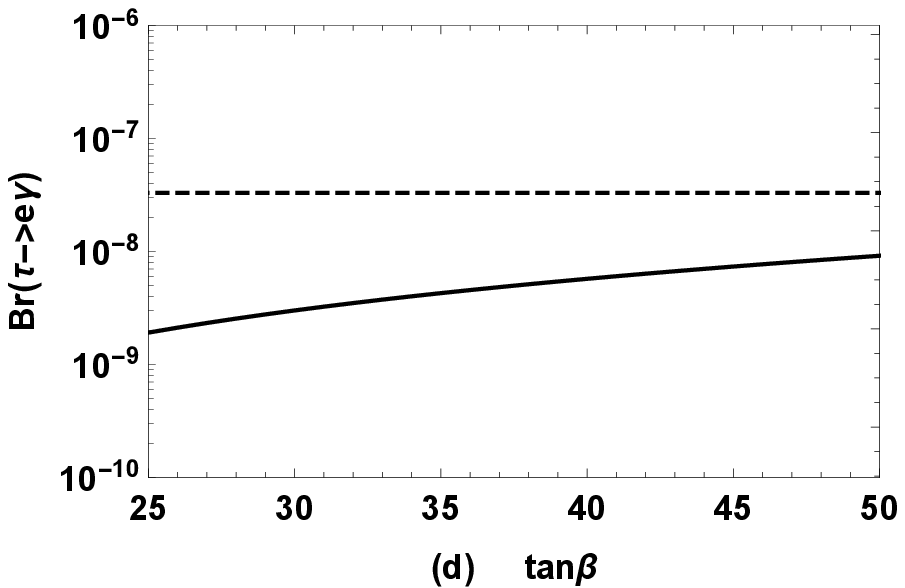}
\vspace{0cm}
\includegraphics[width=2.4in]{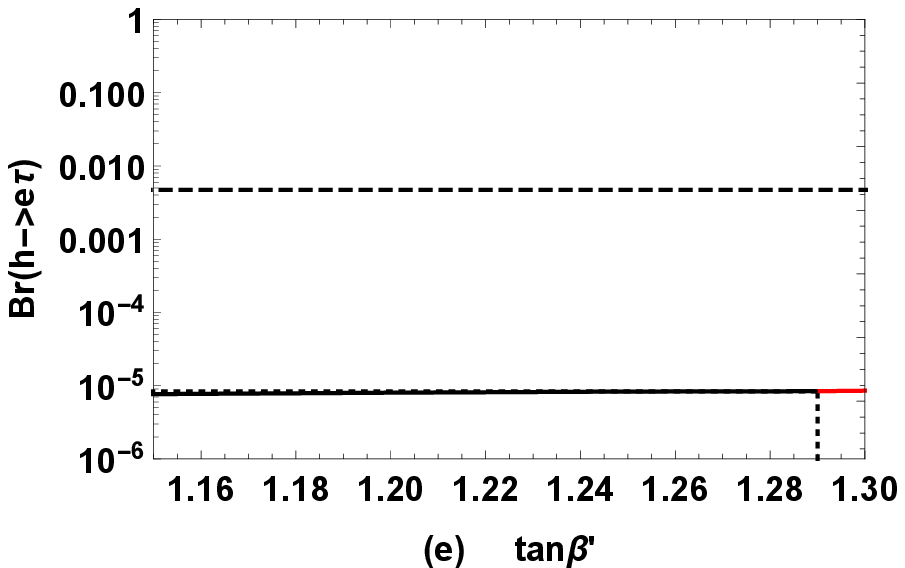}
\vspace{0cm}
\includegraphics[width=2.4in]{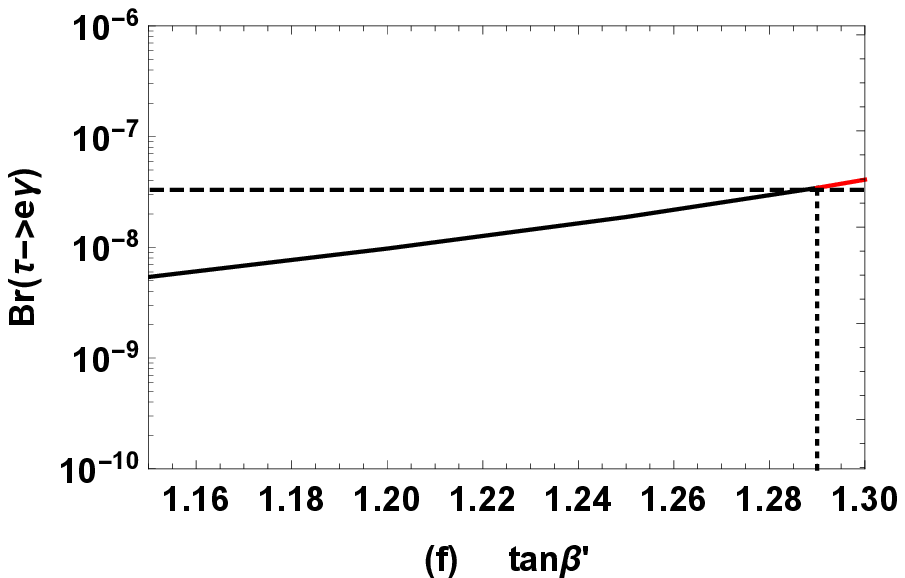}
\vspace{0cm}
\includegraphics[width=2.4in]{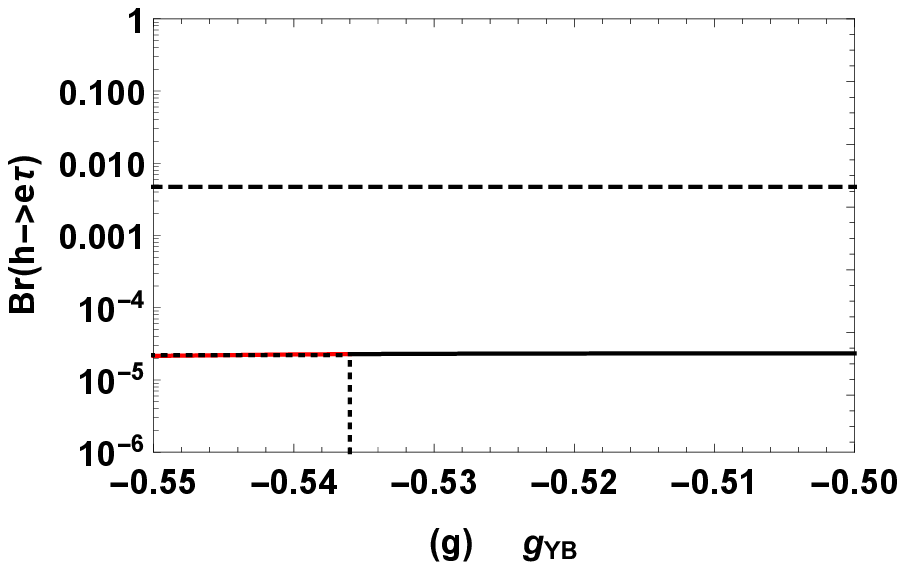}
\vspace{0cm}
\includegraphics[width=2.4in]{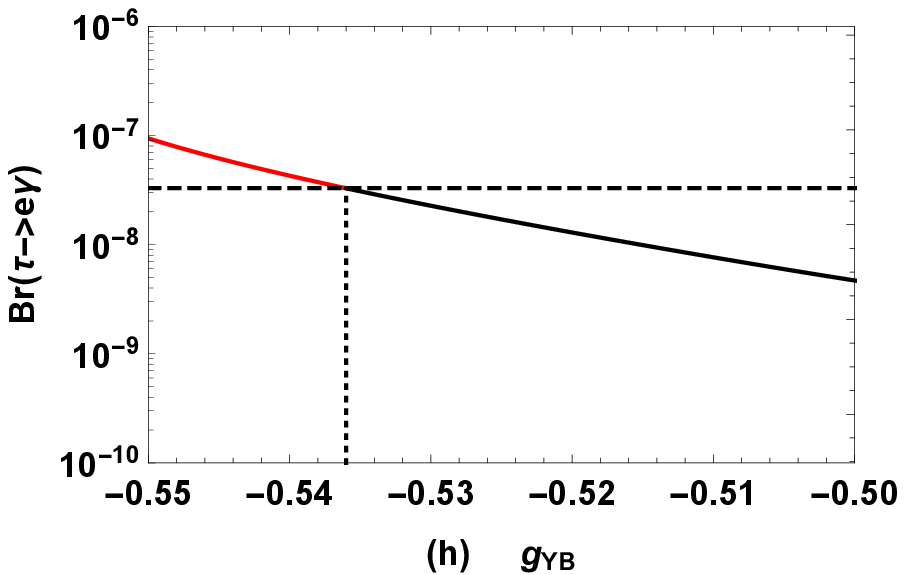}
\vspace{0cm}
\caption[]{{\label{5}} ${\rm{Br}}(h\rightarrow e\tau)$  versus other basic parameters $M_E$ (a), $\tan\beta$ (c), $\tan\beta'$ (e), and $g_{YB}$ (g), where the dashed line stands for the upper limit on ${\rm{Br}}(h\rightarrow e\tau )$ at 95\% C.L. as shown in Eq.(\ref{a2}). ${\rm{Br}}(\tau \rightarrow e\gamma)$ versus other basic parameters $M_E$ (b), $\tan\beta$ (d), $\tan\beta'$ (f), and $g_{YB}$ (h), where the dashed line denotes the present limit of ${\rm{Br}}(\tau\rightarrow e\gamma)$ at 90\% C.L. as shown in Eq.(\ref{a5}). The red solid line is ruled out by the present limit of ${\rm{Br}}(\tau \rightarrow e\gamma)$, and the black solid line is consistent with the present limit of ${\rm{Br}}(\tau \rightarrow e\gamma)$.}
\end{figure}

\begin{figure}
\setlength{\unitlength}{0mm}
\centering
\includegraphics[width=2.6in]{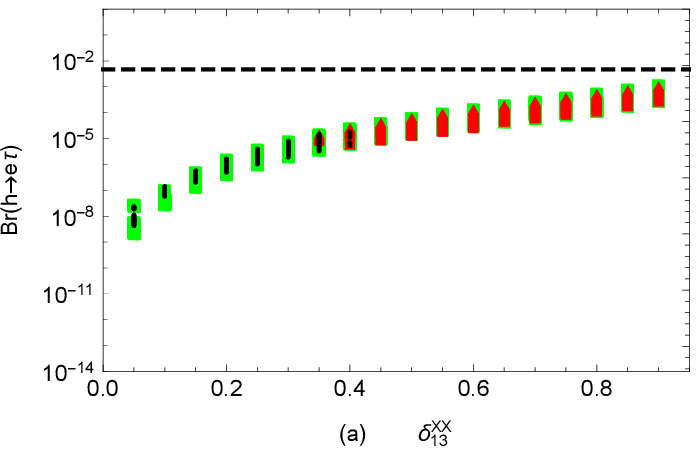}
\vspace{0cm}
\includegraphics[width=2.6in]{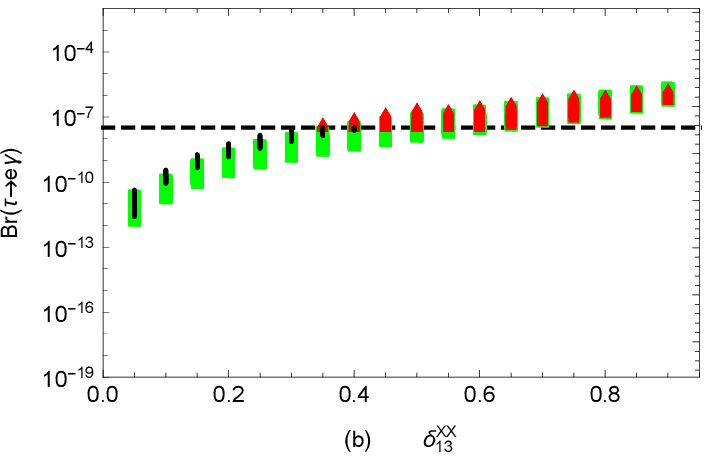}
\vspace{0cm}
\includegraphics[width=2.6in]{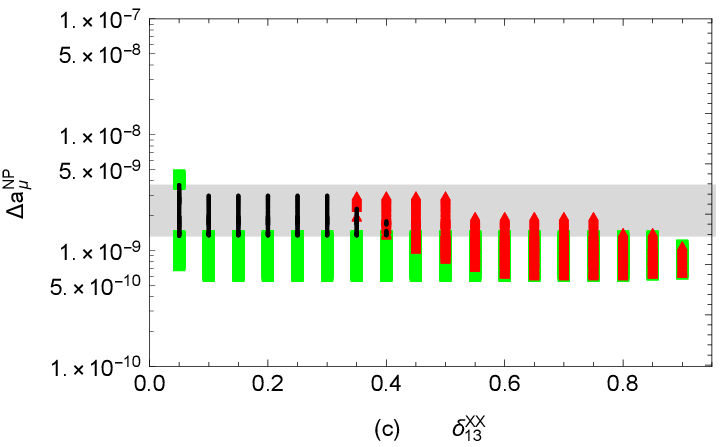}
\vspace{0cm}
\caption[]{{\label{9}} ${\rm{Br}}(h\rightarrow e\tau)$  versus slepton flavor mixing parameters $\delta_{13}^{XX}$ (a) , where the dashed line stands for the upper limit on ${\rm{Br}}(h\rightarrow e\tau )$ at 95\% C.L. as shown in Eq.(\ref{a2}). ${\rm{Br}}(\tau \rightarrow e\gamma)$ versus slepton flavor mixing parameters $\delta_{13}^{XX}$ (b), where the dashed line denotes the present limit of ${\rm{Br}}(\tau \rightarrow e\gamma)$ at 90\% C.L. as shown in Eq.(\ref{a5}). $\Delta a^{NP}_\mu$ versus slepton flavor mixing parameters $\delta_{13}^{XX}$ (c), where the gray area denotes the $\Delta a_\mu$ at $2\sigma$ given in Eq.(\ref{MDM-exp}). Here, the red triangles are excluded by the present limit of  ${\rm{Br}}(\tau \rightarrow e\gamma)$, the green squares are eliminated by the $\Delta a_\mu$ at $2\sigma$, and the black circles simultaneously conform to the present limit of ${\rm{Br}}(\tau \rightarrow e\gamma)$ and the $\Delta a_\mu$ at $2\sigma$.}
\end{figure}

In Fig.~\ref{4}, we can clearly see the influence of slepton mixing parameters $\delta^{XX}_{13}$ $(X=L, R)$ on ${\rm{Br}}(h\rightarrow e\tau) $ and ${\rm{Br}}(\tau \rightarrow e\gamma)$. With the increase of $\delta^{XX}_{13}$ $(X=L, R)$, ${\rm{Br}}(h\rightarrow e\tau) $ and ${\rm{Br}}(\tau \rightarrow e\gamma)$ all increase. ${\rm{Br}}(h\rightarrow e\tau) $ and ${\rm{Br}}(\tau \rightarrow e\gamma) $ are proportional to the slepton mixing parameters $\delta^{XX}_{13}$  $(X=L, R)$. However, the difference is that ${\rm{Br}}(\tau \rightarrow e\gamma) $ can easily attain the experimental upper limit with the increase of $\delta^{XX}_{13}$ $(X=L, R)$, but ${\rm{Br}}(h\rightarrow e\tau) $ cannot do that under the same parameter space. The experimental upper limit of ${\rm{Br}}(h\rightarrow e\tau) $ is  $4.7\times10^{-3}$, which is still insensitive now. In Fig.~\ref{4}, the red solid line is ruled out by the present limit of ${\rm{Br}}(\tau \rightarrow e\gamma)$, and the black solid line is consistent with the present limit of ${\rm{Br}}(\tau \rightarrow e\gamma)$. Considering the upper limit of the experiment and the limit of ${\rm{Br}}(\tau \rightarrow e\gamma) $, ${\rm{Br}}(h\rightarrow e\tau)$ can approximate to $\mathcal{O}(10^{-5})$, which is about only 2 orders of magnitude away from the experimental upper limit. Compared with the MSSM, the B-LSSM adds two new singlet Higgs fields and three generations of right-handed neutrinos, which give new sources for lepton flavor violation and make a  contribution to search for NP.

In order to see the influence of other basic parameters on the numerical results, we again set appropriate numerical values for slepton flavor mixing parameters such as $\delta_{13}^{LL}=0.6$, $\delta_{13}^{RR}=0.4$, and $\delta_{13}^{LR}=0.8$. Then, we look at the influence of basic parameters $M_E$, $\tan\beta$, $\tan\beta'$, and $g_{YB}$ on ${\rm{Br}}(h\rightarrow e\tau)$ and ${\rm{Br}}(\tau \rightarrow e\gamma)$, respectively, which can be intuitively seen in Fig.~\ref{5}. ${\rm{Br}}(h\rightarrow e\tau)$ and ${\rm{Br}}(\tau \rightarrow e\gamma)$ decrease with the increase of $M_E$ or $g_{YB}$, and the branching ratios for these processes $h\rightarrow e\tau$ and $\tau \rightarrow e\gamma$ increase with the increase of $\tan\beta$ or $\tan\beta'$. $M_E$, $\tan\beta$, $\tan\beta'$, and $g_{YB}$ affect the numerical results mainly through the new mass matrix of sleptons, Higgs bosons, and neutralino. Since the two decay processes contain different coupling vertices, the basic parameters $M_E$, $\tan\beta$, $\tan\beta'$, and $g_{YB}$ have different effects on the two processes.

By scanning the parameter space shown in Table~\ref{tab2} with $\delta_{13}^{LL}=\delta_{13}^{RR}=\delta_{13}^{LR}=\delta_{13}^{XX}$, we can clearly see the constraints of ${\rm{Br}}(\tau \rightarrow e\gamma)$ and $\Delta a_\mu$ on ${\rm{Br}}(h\rightarrow e\tau)$ from Fig.~\ref{9}. Although there are many points close to the experimental upper limit in Fig.~\ref{9}(a), only the black circles satisfy the constraints of ${\rm{Br}}(\tau \rightarrow e\gamma)$ and $\Delta a_\mu$. By looking at the black circles in Fig.~\ref{9}(a), ${\rm{Br}}(h\rightarrow e\tau)$ can approach $\mathcal{O}(10^{-5})$, which is about only 2 orders of magnitude away from the experimental upper limit. Perhaps in the near future, the accuracy of the experimental upper limit will be further improved, and the 125 GeV Higgs boson decays with LFV may be detected.

\subsection{125 GeV Higgs boson decays with lepton flavor violation  $h\rightarrow \mu\tau $ }
In the last, we  analyze the process 125 GeV Higgs boson decays with LFV $h\rightarrow \mu\tau$ in the B-LSSM. We still consider the influence of slepton flavor mixing parameters $\delta_{23}^{XX}$ $(X=L, R)$ on ${\rm{Br}}(h\rightarrow \mu\tau) $ and ${\rm{Br}}(\tau \rightarrow \mu\gamma)$ first, and then we consider the restriction of the experimental upper limit of rare process ${\rm{Br}}(\tau \rightarrow \mu\gamma) $ on process ${\rm{Br}}(h\rightarrow \mu\tau) $.

\begin{figure}
\setlength{\unitlength}{1mm}
\centering
\includegraphics[width=2.4in]{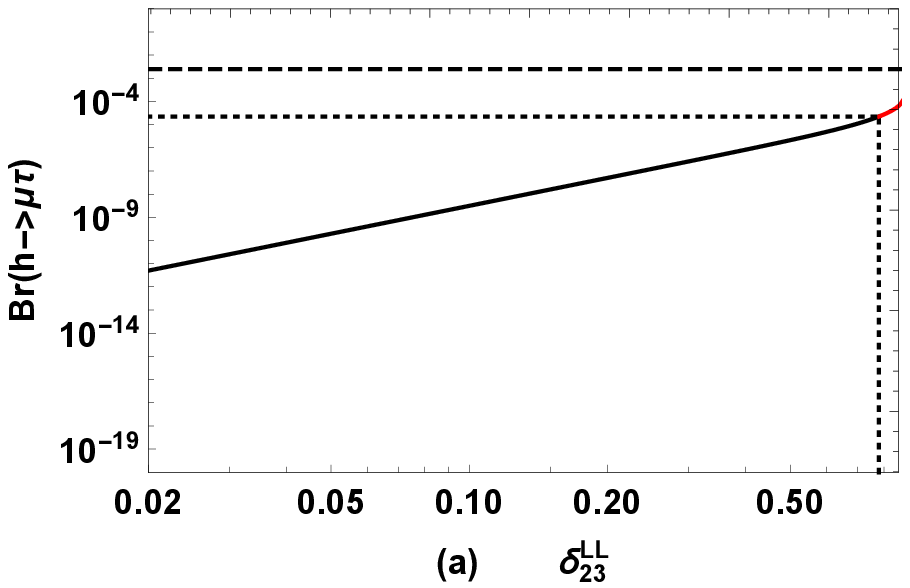}
\vspace{0cm}
\includegraphics[width=2.4in]{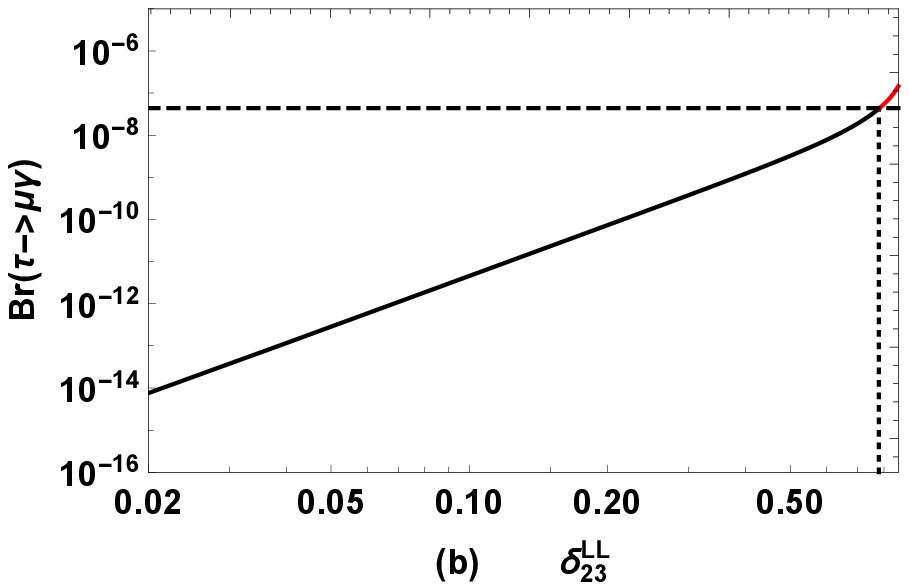}
\vspace{0cm}
\includegraphics[width=2.4in]{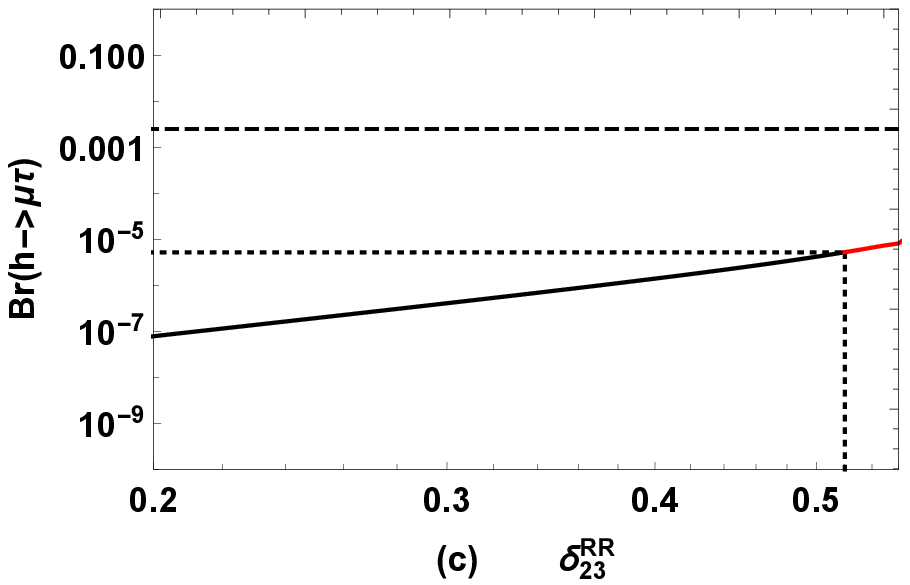}
\vspace{0cm}
\includegraphics[width=2.4in]{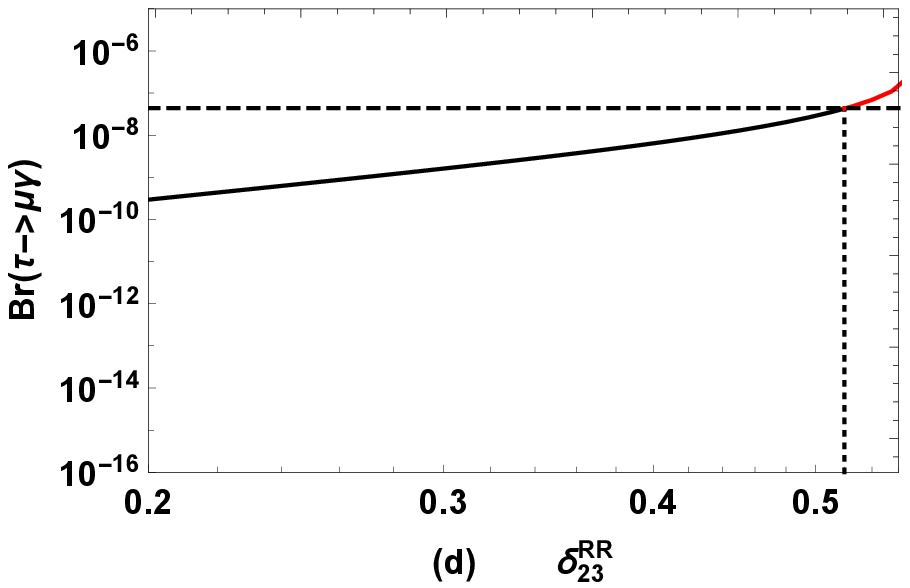}
\vspace{0cm}
\includegraphics[width=2.4in]{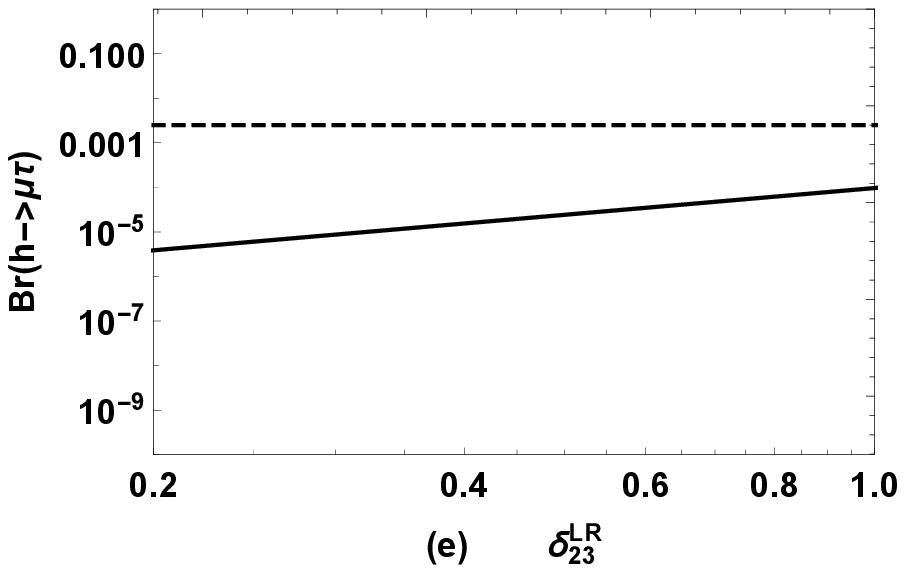}
\vspace{0cm}
\includegraphics[width=2.4in]{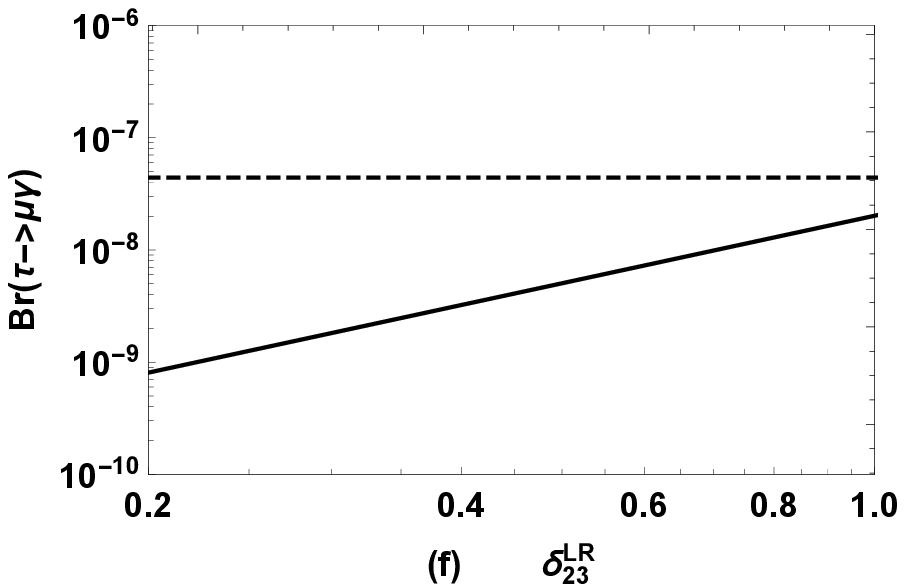}
\vspace{0cm}
\caption[]{{\label{6}} ${\rm{Br}}(h\rightarrow \mu\tau)$  versus slepton flavor mixing parameters $\delta_{23}^{LL}$ (a), $\delta_{23}^{RR}$ (c), and $\delta_{23}^{LR}$ (e), where the dashed line stands for the upper limit on ${\rm{Br}}(h\rightarrow \mu\tau )$ at 95\% C.L. as shown in Eq.(\ref{a3}). ${\rm{Br}}(\tau \rightarrow \mu\gamma)$ versus slepton mixing parameters $\delta_{23}^{LL}$ (b), $\delta_{23}^{RR}$ (d), and $\delta_{23}^{LR}$ (f), where the dashed line denotes the present limit of ${\rm{Br}}(\tau \rightarrow \mu\gamma)$ at 90\% C.L. as shown in Eq.(\ref{a6}). The red solid line is ruled out by the present limit of ${\rm{Br}}(\tau \rightarrow \mu\gamma)$, and the black solid line is consistent with the present limit of ${\rm{Br}}(\tau \rightarrow \mu\gamma)$.}
\end{figure}

\begin{figure}
\begin{center}
\setlength{\unitlength}{2mm}
\centering
\includegraphics[width=2.4in]{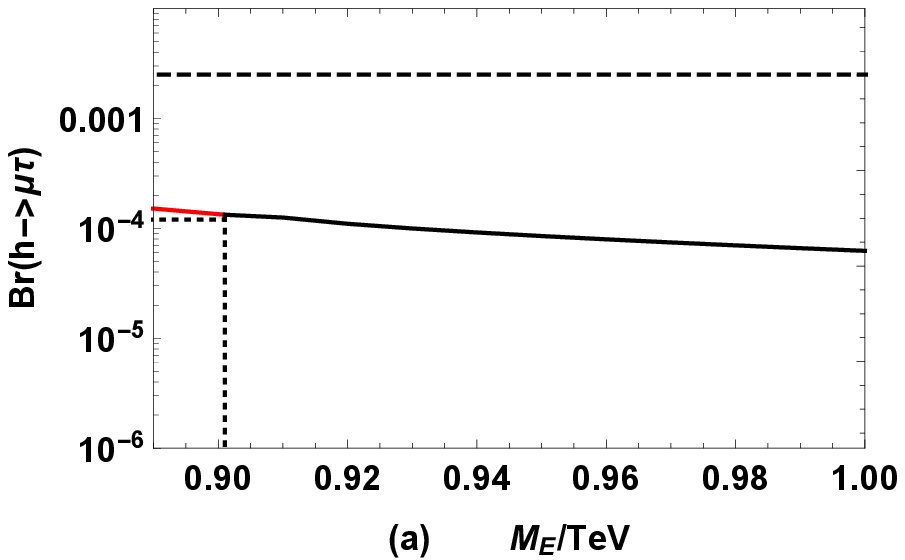}
\vspace{0cm}
\includegraphics[width=2.4in]{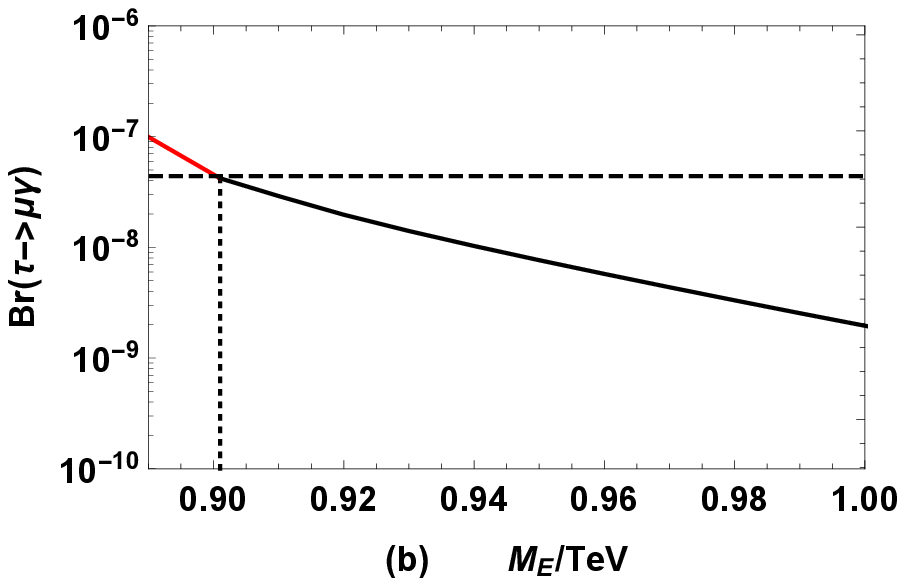}
\vspace{0cm}
\includegraphics[width=2.4in]{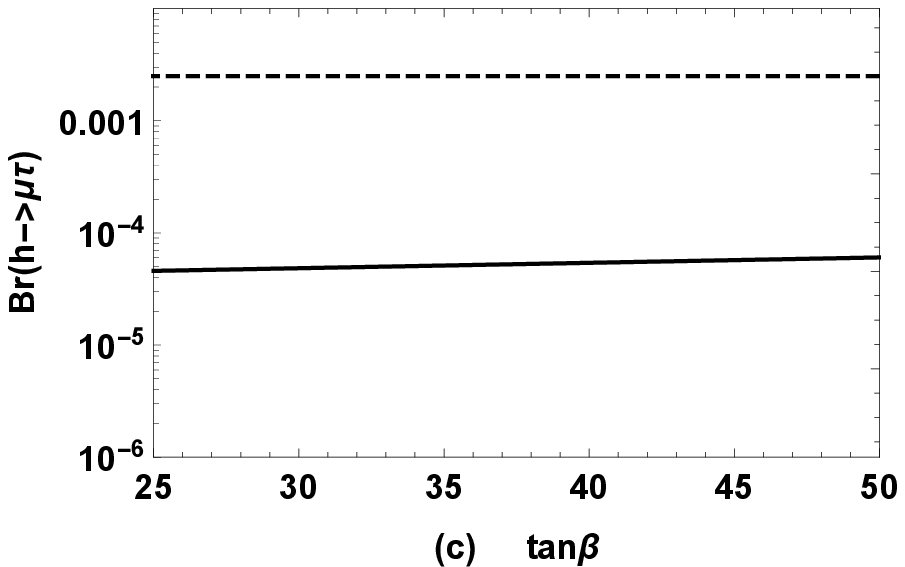}
\vspace{0cm}
\includegraphics[width=2.4in]{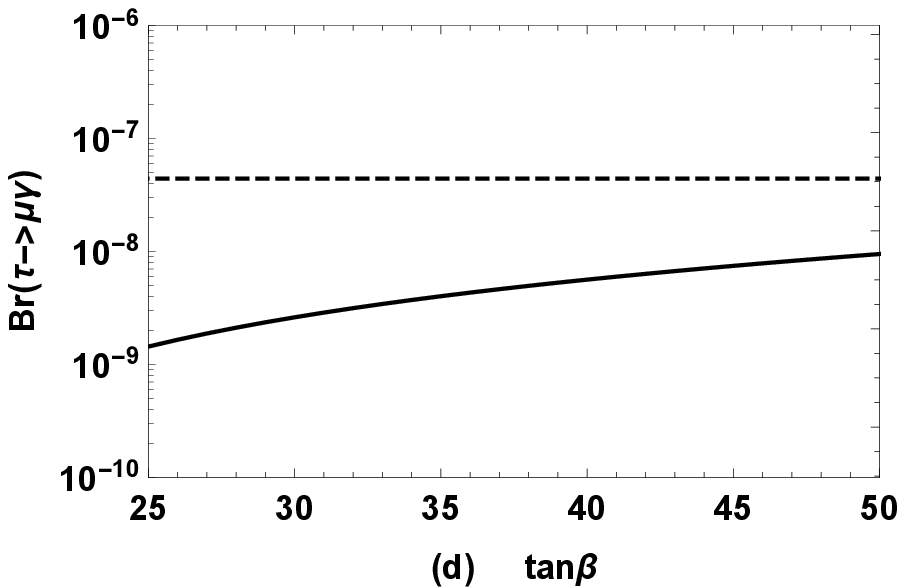}
\vspace{0cm}
\includegraphics[width=2.4in]{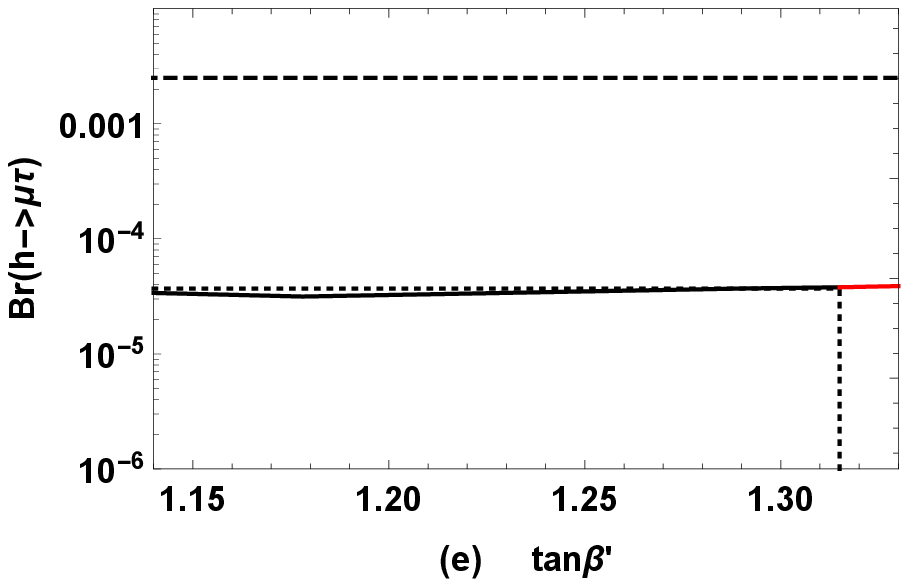}
\vspace{0cm}
\includegraphics[width=2.4in]{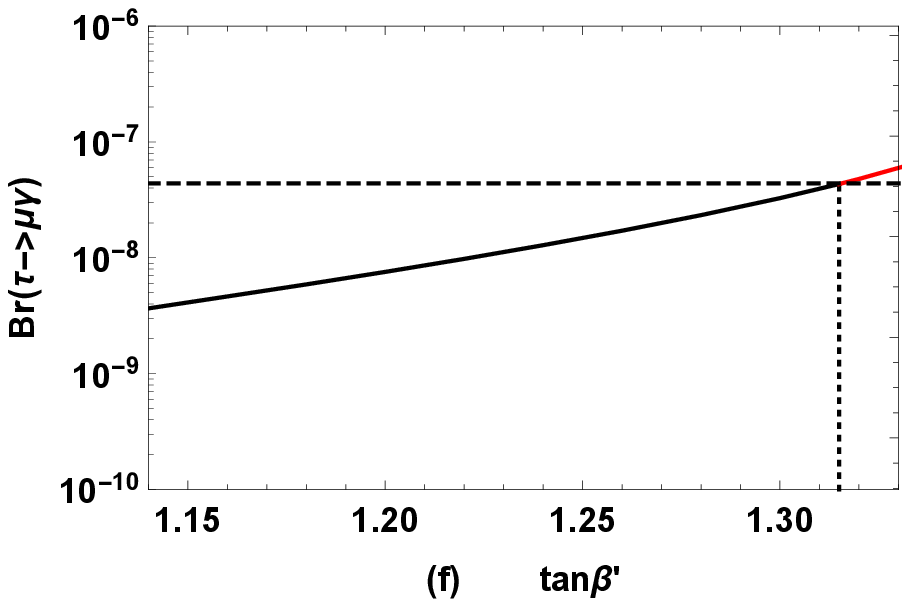}
\vspace{0cm}
\includegraphics[width=2.4in]{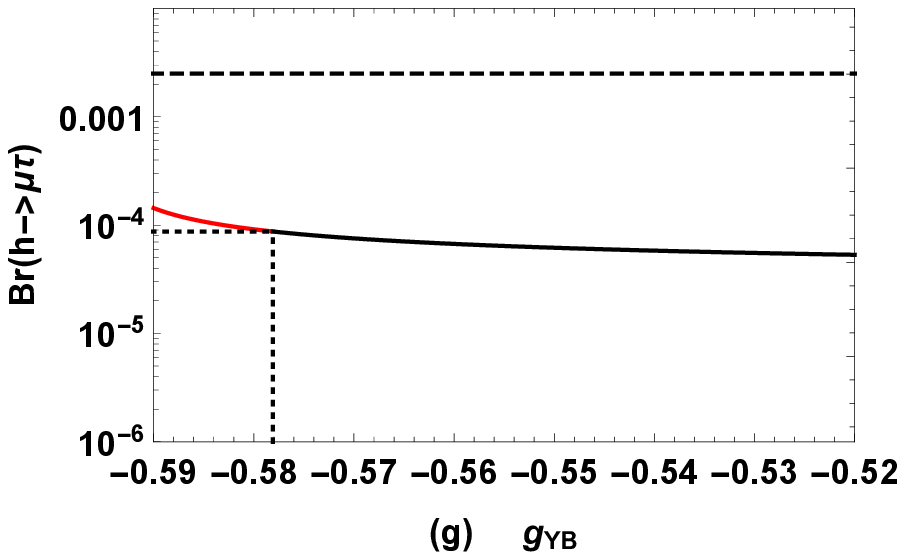}
\vspace{0cm}
\includegraphics[width=2.4in]{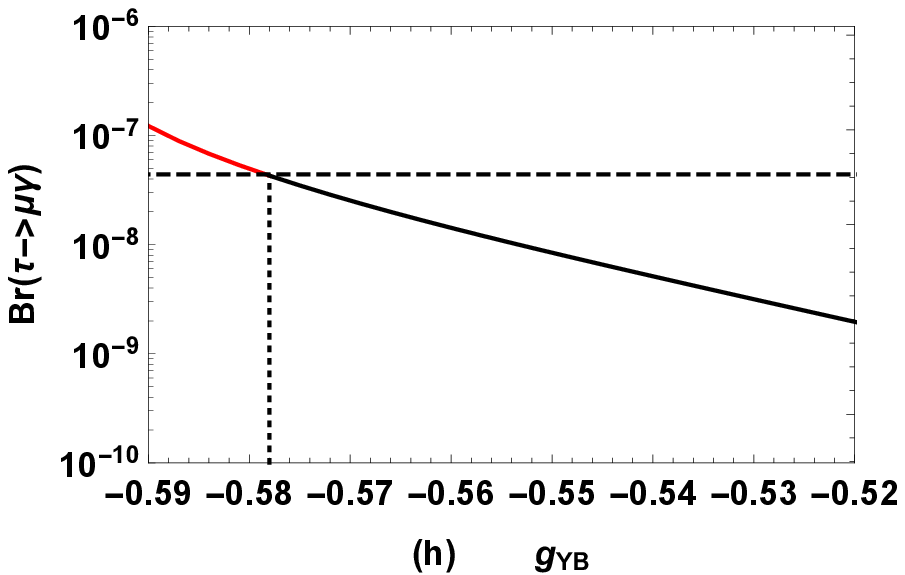}
\vspace{0cm}
\caption[]{{\label{7}} ${\rm{Br}}(h\rightarrow \mu\tau)$  versus other basic parameters $M_E$ (a), $\tan\beta$ (c), $\tan\beta'$ (e), and $g_{YB}$ (g), where the dashed line stands for the upper limit on ${\rm{Br}}(h\rightarrow \mu\tau )$ at 95\% C.L. as shown in Eq.(\ref{a3}). ${\rm{Br}}(\tau \rightarrow \mu\gamma)$ versus other basic parameters $M_E$ (b), $\tan\beta$ (d), $\tan\beta'$ (f), and $g_{YB}$ (h), where the dashed line denotes the present limit of ${\rm{Br}}(\tau\rightarrow \mu\gamma)$ at 90\% C.L. as shown in Eq.(\ref{a6}). The red solid line is ruled out by the present limit of ${\rm{Br}}(\tau \rightarrow \mu\gamma)$, and the black solid line is consistent with the present limit of ${\rm{Br}}(\tau \rightarrow \mu\gamma)$.}
\end{center}
\end{figure}

In Fig.~\ref{6}, we plot the influence of slepton  flavor mixing parameters $\delta^{XX}_{23}$  $(X=L, R)$ on ${\rm{Br}}(h\rightarrow \mu\tau) $ and ${\rm{Br}}(\tau \rightarrow \mu\gamma) $, where the dashed line still represents the experimental upper limit of ${\rm{Br}}(h\rightarrow \mu\tau) $ and ${\rm{Br}}(\tau \rightarrow \mu\gamma) $. The red solid line is ruled out by the present limit of ${\rm{Br}}(\tau \rightarrow \mu\gamma)$, and the black solid line is consistent with the present limit of ${\rm{Br}}(\tau \rightarrow \mu\gamma)$. Figures~\ref{6}(a) and ~\ref{6}(b) show that the branching ratios increase with the increase of $\delta^{LL}_{23}$, and ${\rm{Br}}(\tau \rightarrow \mu\gamma) $ can exceed the experimental upper limit. Although ${\rm{Br}}(h\rightarrow \mu\tau) $ fails to reach the experimental upper limit, it is close to the experimental upper limit of ${\rm{Br}}(h\rightarrow \mu\tau) $. In Figs.~\ref{6}(c) and ~\ref{6}(d), the branching ratios also increase along with the growth of $\delta^{RR}_{23}$, where ${\rm{Br}}(\tau \rightarrow \mu\gamma) $ can still reach the experimental upper limit, but ${\rm{Br}}(h\rightarrow \mu\tau) $ cannot do that. Figures~\ref{6}(e) and ~\ref{6}(f) show that ${\rm{Br}}(h\rightarrow \mu\tau) $ and ${\rm{Br}}(\tau \rightarrow \mu\gamma) $ are all very close to the experimental upper limits but below the upper limit of the experiment. ${\rm{Br}}(h\rightarrow \mu\tau) $ can be up to $\mathcal{O}(10^{-4})$. Since the LFV processes are flavor dependent, the slepton flavor mixing parameters $\delta^{XX}_{23}$  $(X=L, R)$ have a large influence on ${\rm{Br}}(h\rightarrow \mu\tau) $ and ${\rm{Br}}(\tau \rightarrow \mu\gamma) $.

It is the same as in the previous analysis in order to see the influence of other basic parameters on the numerical results. We again set appropriate numerical values for slepton flavor mixing parameters such as $\delta_{23}^{LL}=0.6$, $\delta_{23}^{RR}=0.4$, and $\delta_{23}^{LR}=1$. In Fig.~\ref{7}, we picture the influence of the four basic parameters of $M_E$, $\tan\beta$, $\tan\beta'$, and $g_{YB}$ on ${\rm{Br}}(h\rightarrow \mu\tau)$ and ${\rm{Br}}(\tau \rightarrow \mu\gamma)$. The branching ratios decrease with the increase of mass $M_E$, and $M_E$ affects the numerical results mainly through the mass of sleptons. When $M_E$ is small, ${\rm{Br}}(\tau \rightarrow \mu\gamma)$ can attain the experimental upper limit, but ${\rm{Br}}(h\rightarrow \mu\tau)$ is still under the experimental upper limit. The branching ratios increase with the increase of $\tan\beta$ or $\tan\beta'$. $\tan\beta$ and $\tan\beta'$ affect the numerical results mainly through the new mass matrix of sleptons, Higgs bosons, and neutralino. The branching ratios decrease with the increase of the parameter $g_{YB}$. $g_{YB}$ is a new parameter in B-LSSM; it affects the numerical results mainly through the new mass matrix of sleptons. Constrained by the experimental upper limit of ${\rm{Br}}(\tau \rightarrow \mu\gamma)$, ${\rm{Br}}(h\rightarrow \mu\tau) $ in the B-LSSM can reach $\mathcal{O}(10^{-4})$, which can be easily seen in Figs.~\ref{7}(a)-\ref{7}(g).

\begin{figure}
\setlength{\unitlength}{0mm}
\centering
\includegraphics[width=2.6in]{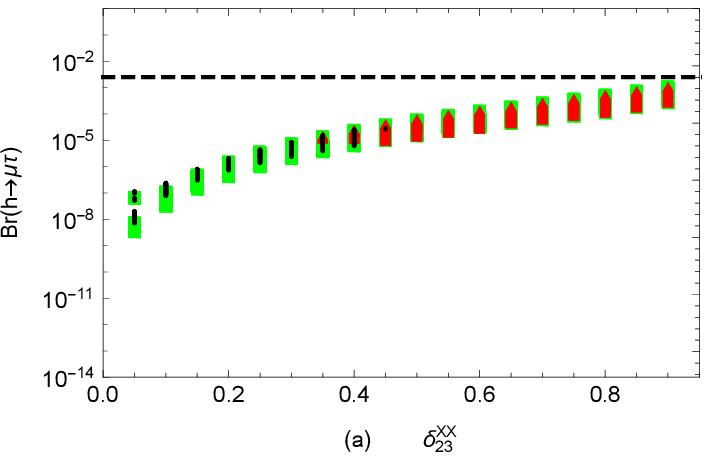}
\vspace{0cm}
\includegraphics[width=2.6in]{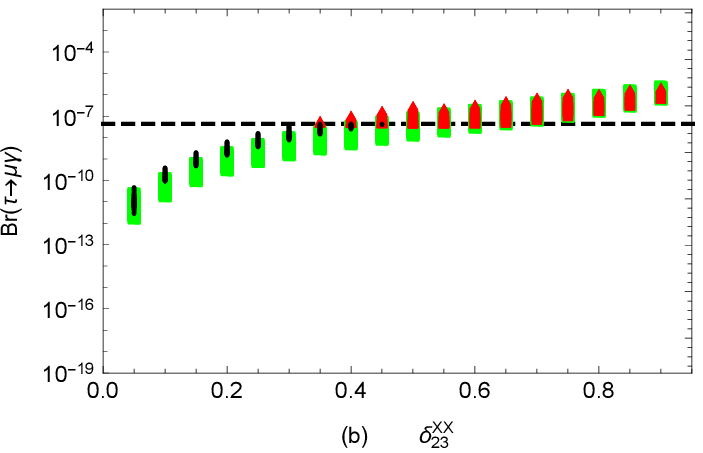}
\vspace{0cm}
\includegraphics[width=2.6in]{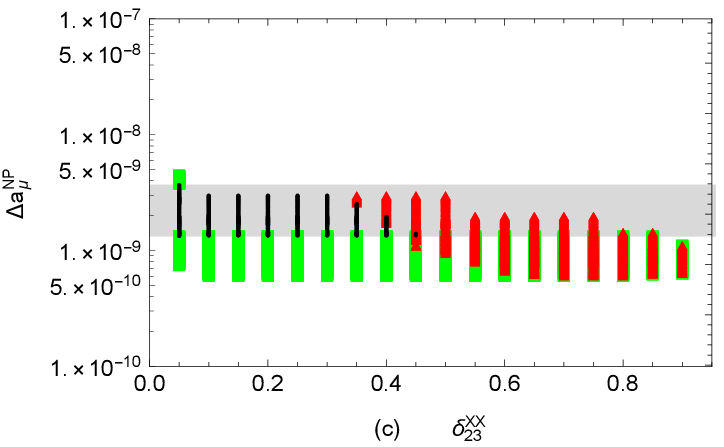}
\vspace{0cm}
\caption[]{{\label{10}} ${\rm{Br}}(h\rightarrow \mu\tau)$  versus slepton flavor mixing parameters $\delta_{23}^{XX}$ (a) , where the dashed line stands for the upper limit on ${\rm{Br}}(h\rightarrow \mu\tau )$ at 95\% C.L. as shown in Eq.(\ref{a3}). ${\rm{Br}}(\tau\rightarrow \mu\gamma)$ versus slepton flavor mixing parameters $\delta_{23}^{XX}$ (b), where the dashed line denotes the present limit of ${\rm{Br}}(\tau\rightarrow \mu\gamma)$ at 90\% C.L. as shown in Eq.(\ref{a6}). $\Delta a^{NP}_\mu$ versus slepton flavor mixing parameters $\delta_{23}^{XX}$ (c), where the gray area denotes the $\Delta a_\mu$ at $2\sigma$ given in Eq.(\ref{MDM-exp}). Here, the red triangles are excluded by the present limit of  ${\rm{Br}}(\tau\rightarrow \mu\gamma)$, the green squares are eliminated by the $\Delta a_\mu$ at $2\sigma$, and the black circles simultaneously conform to the present limit of ${\rm{Br}}(\tau\rightarrow \mu\gamma)$ and the $\Delta a_\mu$ at $2\sigma$.}
\end{figure}

By scanning the parameter space shown in Table~\ref{tab2}, we also get the related images of ${\rm{Br}}(h\rightarrow \mu\tau)$, ${\rm{Br}}(\tau \rightarrow \mu\gamma)$, $\Delta a^{NP}_\mu$  versus  $\delta_{23}^{XX}$ (here $\delta_{23}^{LL}=\delta_{23}^{RR}=\delta_{23}^{LR}=\delta_{23}^{XX}$) as shown in Fig.~\ref{10}. After taking into account the constraints of ${\rm{Br}}(\tau \rightarrow \mu\gamma)$ and $\Delta a_\mu$ on ${\rm{Br}}(h\rightarrow \mu\tau)$,  ${\rm{Br}}(h\rightarrow \mu\tau) $ in the B-LSSM can approximate to $\mathcal{O}(10^{-4})$. In the near future, the 125 GeV Higgs boson decays with LFV $h\rightarrow \mu\tau$ may be detected, by improving the accuracy of the experimental upper limit.

\section{Summary\label{sec5}}
In this work, we have studied the 125 GeV Higgs boson decays with lepton flavor violation, $h\rightarrow e\mu$, $h\rightarrow e\tau$, and $h\rightarrow \mu\tau$, in the framework of the B-LSSM with slepton flavor mixing. The numerical results show that the branching ratios of $h\rightarrow e\mu$, $h\rightarrow e\tau$, and $h\rightarrow \mu\tau$ depend on the slepton flavor mixing parameters $\delta_{12}^{XX}~(X=L,R)$, $\delta_{13}^{XX}~(X=L,R)$, and $\delta_{23}^{XX}~(X=L,R)$, respectively, because the lepton flavour violating  processes are flavor dependent. Under the experimental constraints of the LFV rare decays $\mu\rightarrow e\gamma$, $\tau\rightarrow e\gamma$, $\tau\rightarrow \mu\gamma$, and muon $(g-2)$, the branching ratio of $h\rightarrow e\mu$ can reach $\mathcal{O}(10^{-11})$, the branching ratio of $h\rightarrow e\tau$ can come up to $\mathcal{O}(10^{-5})$, and the branching ratio of $h\rightarrow \mu \tau$ can approach $\mathcal{O}(10^{-4})$, respectively. The branching ratios of $h\rightarrow e\tau$ and  $h\rightarrow \mu \tau$ in the B-LSSM are close to the experimental upper limits of ${\rm{Br}}(h\rightarrow e\tau) $ and ${\rm{Br}}(h\rightarrow \mu\tau) $, which may be detected in the future. Compared with the MSSM, exotic two singlet Higgs fields and three generations of right-handed neutrinos in the B-LSSM induce new sources for the lepton flavor violation. $\tan\beta'$ and $g_{YB}$ are new parameters in the B-LSSM, which can affect the numerical results through the new mass matrix of sleptons, Higgs bosons, and neutralino. Considering that the recent ATLAS and CMS measurements for $h\rightarrow e\mu$, $h\rightarrow e\tau$, and $h\rightarrow \mu\tau$ do not show a significant deviation from the SM, the experiments still need to make more precise measurements in the future. To detect the Higgs boson lepton flavor violating process is a prospective window to search for new physics.

\begin{acknowledgments}
\indent\indent
The work was supported by the National Natural Science Foundation of China with Grants No. 11705045, No. 12075074, and No. 11535002, Natural Science Foundation of Hebei province with Grant No. A2020201002, the youth top-notch talent support program of the Hebei Province, and Midwest Universities Comprehensive Strength Promotion project.
\end{acknowledgments}

\appendix

\section{loop function \label{loop function}}
The loop function $G$ is written by
\begin{eqnarray}
&&\hspace{-0.75cm}{G_1}({\textit{x}_1 , \textit{x}_2 , \textit{x}_3}) =  \frac{1}{{16{\pi ^2}}}\Big[ \frac{{{x_1}\ln {x_1}}}{{({x_2} - {x_1})({x_1} - {x_3})}}
+ \frac{{{x_2}\ln {x_2}}}{{({x_1} - {x_2})({x_2} - {x_3})}}  + \frac{{{x_3}\ln {x_3}}}{{({x_1} - {x_3})({x_3} - {x_2})}}\Big], \\
&&\hspace{-0.75cm}{G_2}({\textit{x}_1 , \textit{x}_2 , \textit{x}_3}) =  \frac{1}{{16{\pi ^2}}}\Big[  \frac{{x_1^2\ln {x_1}}}{{({x_2} - {x_1})({x_1} - {x_3})}}
+ \frac{{x_2^2\ln {x_2}}}{{({x_1} - {x_2})({x_2} - {x_3})}}  + \frac{{x_3^2\ln {x_3}}}{{({x_1} - {x_3})({x_3} - {x_2})}} \Big].
\end{eqnarray}

\section{The couplings \label{app-coupling}}

The coupling of neutral scalars and charged scalars can be written as
\begin{eqnarray}
&&C^{S^c}_{1mn}=1/4(-2(\sqrt{2}\sum_{b=1}^3Z_{mb}^{E,*}\sum_{a=1}^3Z_{n(3+a)}^ET_{e,ab}Z_{k1}^H+\sqrt{2}\sum_{b=1}^3\sum_{a=1}^3Z_{m(3+a)}^{E,*}T_{e,ab}^*Z_{nb}^EZ_{k1}^H\nonumber\\
&&\hspace{0.9cm}+2v_d\sum_{c=1}^3Z_{m(3+c)}^{E,*}\sum_{b=1}^3\sum_{a=1}^3Y_{e,ca}^*Y_{e,ba}Z_{k(3+b)}^EZ_{k1}^H+2v_d\sum_{c=1}^3\sum_{b=1}^3Z_{mb}^{E,*}\sum_{a=1}^3Y_{e,ac}^*Y_{e,ab}Z_{nc}^EZ_{k1}^H\nonumber\\
&&\hspace{0.9cm}-\sqrt{2}\mu^*\sum_{b=1}^3Z_{mb}^{E,*}\sum_{a=1}^3Y_{e,ab}Z_{n(3+a)}^EZ_{k2}^H-\sqrt2 \mu \sum_{b=1}^3\sum_{a=1}^3Y_{e,ab}^*Z_{m(3+a)}^{E,*}Z_{nb}^EZ_{k2}^H)\nonumber\\
&&\hspace{0.9cm}+\sum_{a=1}^3Z_{m(3+a)}^{E,*}Z_{n(3+a)}^E((2g_1^2+g_{YB}(2g_{YB}+g_B))v_dZ_{k1}^H-(2g_1^2+g_{YB}(2g_{YB}+g_B))v_uZ_{k2}^H\nonumber\\
&&\hspace{0.9cm}+2(2g_{YB}g_B+g_B^2)(-v_{\bar\eta}Z_{k4}^H+v_\eta Z_{k3}^H))-\sum_{a=1}^3Z_{ma}^{E,*}Z_{na}^E((-g_2^2+g_{YB}g_B+g_1^2+g_{YB}^2)v_dZ_{k1}^H\nonumber\\
&&\hspace{0.9cm}-(-g_2^2+g_{YB}g_B+g_1^2+g_{YB}^2)v_uZ_{k2}^H+2(g_{YB}g_B+g_B^2)(-v_{\bar \eta} Z_{k4}+v_\eta Z_{k3}^H))).
\end{eqnarray}

The couplings of charged scalars, neutral fermions and charged fermions are written as
\begin{eqnarray}
&&C_L^{S_n^c \chi _\eta ^o{{\bar l }_{i}}}=-\sqrt2g_1N_{\eta1}^*\sum_{a=1}^3Z_{n(3+a)}^{E,*}U_{R,ia}^{e,*}
-1/\sqrt2(2g_{YB}+g_B)N_{\eta5}^*\sum_{a=1}^3Z_{n(3+a)}^{E,*}U_{R,ia}^{e,*}\nonumber\\
&&\hspace{1.5cm}-N_{\eta3}^*\sum_{b=1}^3Z_{nb}^{E,*}\sum_{a=1}^3U_{R,ia}^{e,*}Y_{e,ab},\\
&&C_R^{S_n^c \chi _\eta ^o{{\bar l }_{i}}}=1/2(-2\sum_{b=1}^3\sum_{a=1}^3Y_{e,ab}^*Z_{n(3+a)}^{E,*}U_{L,ib}^eN_{\eta3}+\sqrt2\sum_{a=1}^3Z_{na}^{E,*}U_{L,ia}^e(g_1N_{\eta1}+g_2N_{\eta2}\nonumber\\
&&\hspace{1.5cm}+(g_{YB}+g_B)N_{\eta5})),\\
&&C_L^{S_m^{c*} {l_{j}}\bar \chi _\eta ^o}=1/2(\sqrt2g_1N_{\eta1}^*\sum_{a=1}^3U_{L,ja}^{e,*}Z_{ma}^E+\sqrt2g_2N_{\eta2}^*\sum_{a=1}^3U_{L,ja}^{e,*}Z_{ma}^E
+\sqrt2g_{YB}N_{\eta5}^*\sum_{a=1}^3U_{L,ja}^{e,*}Z_{ma}^E\nonumber\\
&&\hspace{1.7cm}+\sqrt2g_BY_{\eta5}^*\sum_{a=1}^3U_{L,ja}^{e,*}Z_{ma}^E-2N_{\eta3}^*\sum_{b=1}^3U_{L,jb}^{e,*}\sum_{a=1}^3Y_{e,ab}Z_{m(3+a)}^E),\\
&&C_R^{S_m^{c*} {l_{j}}\bar \chi _\eta ^o}=-1/\sqrt2\sum_{a=1}^3Z_{m(3+a)}^EU_{R,ja}^e(2g_1N_{\eta1}+(2g_{YB}+g_B)N_{\eta5})-\sum_{b=1}^3\sum_{a=1}^3Y_{e,ab}^*U_{R,ja}^eZ_{mb}^EN_{\eta3}.\nonumber\\
&&\hspace{1.7cm} \;\;\;\quad
\end{eqnarray}

The couplings of neutral scalars and neutral fermions can be written by
\begin{eqnarray}
&&C_L^{h{\chi _\eta^o }{{\bar \chi }_\sigma^o}}=1/2(-g_2N_{\eta2}^*N_{\sigma3}^*Z^H_{k1}+g_{YB}N_{\eta5}^*N_{\sigma3}^*Z^H_{k1}+N_{\eta3}^*(g_1N_{\sigma1}^*-g_2N_{\sigma2}^*+g_{YB}N_{\sigma5}^*)Z_{k1}^H\nonumber\\
&&\hspace{1.5cm}-g_1N_{\eta4}^*N_{\sigma1}^*Z_{k2}^H+g_2N_{\eta4}^*N_{\sigma2}^*Z_{k2}^H+g_2N_{\eta2}^*N_{\sigma4}^*Z_{k2}^H-g_{YB}N_{\eta5}^*N_{\sigma4}^*Z_{k2}^H\nonumber\\
&&\hspace{1.5cm}-g_{YB}N_{\eta4}^*N_{\sigma5}^*Z_{k2}^H+2g_BN_{\eta6}^*N_{\sigma5}^*Z_{k3}^H+2g_BN_{\eta5}^*N_{\sigma6}^*Z_{k3}^H-2g_BN_{\eta7}^*N_{\sigma5}^*Z_{k4}^H\nonumber\\
&&\hspace{1.5cm}-2g_BN_{\eta5}^*N_{\sigma7}^*Z_{k4}^H+N_{\eta1}^*(g_1N_{\sigma3}^*Z_{k1}^H-g_1N_{\sigma4}^*Z_{k2}^H)),\\
&&C_R^{h{\chi _\eta^o }{{\bar \chi }_\sigma^o}}=1/2(Z_{k1}^H((g_1N_{\eta1}-g_2N_{\eta2}+g_{YB}N_{\eta5})N_{\sigma3}+N_{\eta3}(g_1N_{\sigma1}-g_2N_{\sigma2}+g_{YB}N_{\sigma5}))\nonumber\\
&&\hspace{1.5cm}-Z_{k2}^H((g_1N_{\eta1}-g_2N_{\eta2}+g_{YB}N_{\eta5})N_{\sigma4}+N_{\eta4}(g_1N_{\sigma1}-g_2N_{\sigma2}+g_{YB}N_{\sigma5}))\nonumber\\
&&\hspace{1.5cm}+2(Z_{k3}^H(g_BN_{\eta5}N_{\sigma6}+N_{\eta6}g_BN_{\sigma5})-Z_{k4}^H(g_BN_{\eta5}N_{\sigma7}+N_{\eta7}g_BN_{\sigma5}))).
\end{eqnarray}

The couplings of neutral scalars and charged fermions are written as
\begin{eqnarray}
&&\hspace{-1.3cm}C_L^{h l_i {\bar l}_i}=-1/\sqrt2\sum_{b=1}^3U_{L,ib}^{e,*}\sum_{a=1}^3U_{R,ia}^{e,*}Y_{e,ab}Z_{k1}^H,\\
&&\hspace{-1.3cm}C_R^{h l_i {\bar l}_i}=-1/\sqrt2\sum_{b=1}^3\sum_{a=1}^3Y_{e,ab}^*U_{R,ia}^eU_{L,ib}^eZ_{k1}^H.
\end{eqnarray}
The matrices $Z$, $N$, $U$, and $V$ above can be found in the version of the B-LSSM that is encoded in SARAH~\cite{164,165,166,167,168}.

\end{document}